\documentclass[twocolumn, twocolappendix]{aastex701}

\usepackage{amsmath}
\usepackage{apjfonts}
\usepackage{bm}

\usepackage{tipa}

\newcommand{\schwa}{\textschwa{}}
\newcommand{\eng}{\textipa{N}}
\newcommand{\Lekwnen}{L\schwa{}$\overset{\displaystyle\textrm{,}}{\textrm{k}}$\textsuperscript{w}\schwa{}\eng{}\schwa{}n}

\newcommand{\V}{\textrm{v}}
\newcommand{\vmax}{\V_\text{max}}

\newcommand{\rmax}{r_\text{max}}
\newcommand{\kpc}{\text{kpc}}
\newcommand{\Gyr}{\text{Gyr}}
\newcommand{\kms}{\mathrm{km\,s^{-1}}}
\newcommand{\masyr}{\mathrm{mas\,yr^{-1}}}
\newcommand{\Mo}{\mathrm{M}_\sun}

\newcommand{\agama}{\texttt{Agama}}
\newcommand{\gadget}{\texttt{Gadget-4}}

\defcitealias{jensen+2024}{J+24}
\newcommand{\jj}{\citetalias{jensen+2024}}

\makeatletter
\DeclareRobustCommand\citepos
  {\begingroup
   \let\NAT@nmfmt\NAT@posfmt
   \NAT@swafalse\let\NAT@ctype\z@\NAT@partrue
   \@ifstar{\NAT@fulltrue\NAT@citetp}{\NAT@fullfalse\NAT@citetp}}

\let\NAT@orig@nmfmt\NAT@nmfmt
\def\NAT@posfmt#1{\NAT@orig@nmfmt{#1's}}

\makeatother

\received{April 25, 2026}
\revised{\today}
\submitjournal{ApJ}

\begin{document}

\title{Galactic tides and the outer density profile of the Sculptor and Ursa Minor dwarf spheroidals}
\shorttitle{Galactic tides on Scl and UMi}
\shortauthors{D. A. Boyea et al.}

\author[orcid=0009-0008-8903-160X, gname='Daniel', sname='Boyea']{Daniel A. Boyea}
\affiliation{Department of Physics and Astronomy, University of Victoria, PO Box 1700 STN CSC, Victoria, BC  V8W 2Y2, Canada}
\email[show]{danielaboyea@gmail.com}  
\correspondingauthor{Daniel A. Boyea}

\author[0000-0003-3862-5076, gname='Julio', sname='Navarro']{Julio F. Navarro}
\affiliation{Department of Physics and Astronomy, University of Victoria, PO Box 1700 STN CSC, Victoria, BC  V8W 2Y2, Canada}
\affiliation{KU-KIST Green School, Korea University}
\email[]{jfn@uvic.ca}

\author[0000-0002-4350-7632, gname='Jaclyn', sname='Jensen']{Jaclyn Jensen}
\affiliation{Department of Physics and Astronomy, Dartmouth College, 6127 Wilder Laboratory,  Hanover, NH 03755-3528, USA}
\affiliation{Department of Physics and Astronomy, University of Victoria, PO Box 1700 STN CSC, Victoria, BC  V8W 2Y2, Canada}
\email[]{jaclyn.r.jensen@dartmouth.edu}

\author[]{Rapha\"el Errani}
\affiliation{McWilliams Center for Cosmology and Astrophysics, Department of Physics,
Carnegie Mellon University, Pittsburgh, PA 15213, USA
}
\email[]{errani@cmu.edu}


\begin{abstract}
   Most dwarf spheroidal (dSph) satellites  of the Milky Way follow exponential surface density profiles that decline sharply in the outer regions. The Sculptor (Scl) and Ursa Minor (UMi) dSphs deviate from this trend and show a clear excess of stars in the outskirts. Individual members have recently been identified as far as ${\sim}10$ effective radii from the center in both systems.
   We study whether far-outlying stars in Scl and UMi may result from Galactic tidal forces using idealized N-body simulations.
   Our results indicate that, on their current orbits, neither galaxy has experienced tidal forces strong enough to affect its stellar density profile. The observed velocity dispersion and size of Scl and UMi imply that these dwarfs are simply too dense to have been affected by Galactic tides. Our conclusions do not change when including the effects of the Large Magellanic Cloud, which, according to our simulations, has substantially perturbed Scl's orbit during a recent, close encounter. Our results are insensitive to assumptions about the detailed dark matter density profile of either galaxy, including the presence of an inner core. 
  We conclude that the outlying stars in Scl or UMi are not of tidal origin, but rather innate features that possibly reflect past merger events or the presence of multiple dynamical components.
\end{abstract}



\section{Introduction\label{sec:intro}}

The  structure of a dwarf galaxy holds clues to its
evolutionary history.
Most nearby dwarf spheroidals (dSph) are well-described by an exponential
surface density profile, with a sharp decline in the outer regions \citep[e.g.,][]{mateo1998, mcconnachie+irwin2006, munoz+2018}. 
However, a number of dwarfs show an  excess of distant stars relative to an exponential.
For example, \citet{chiti+2021} find an extended distribution of stars around Tucana II, which they suggested arose from a past merger. In addition, the Sagittarius, Tucana III, Crater II, and Antlia II dSphs show clear evidence of tidal disruption, with streams that extend far from the center of each dwarf \citep[e.g.,][]{ibata+gilmore+irwin1994, li+2018, vivas+2022, ji2025}. 

Among classical dwarf galaxies (i.e., those with $M_V<-8$, or $M_*>10^5\, \Mo$),  Sculptor (Scl)  and Ursa Minor (UMi) display a significant overabundance of distant stars (see Section~\ref{sec:gaia}). \citet{innanen+papp1979} first noted that RR Lyrae candidate members in Scl appeared to extend out to $180'$, or well over 15 half-light radii ($R_h = 9.79' = 240\,\text{pc}$).\footnote{We use sphericalized (half-light) radii throughout this work, see equation~\ref{eq:r_ell} below.}
Subsequent density profile measurements have repeatedly reported  an apparent outer density excess \citep[e.g.,][]{eskridge1988, palma+2003, westfall+2006, sato+2025b}. 
More recently, \citet{sestito+2023a, sestito+2023b} followed up several stellar candidates from \cite{jensen+2024} and spectroscopically confirmed members out to ${\sim}10$ half-light radii ($R_h$) from the center of Scl and UMi. If these galaxies were well described by exponential profiles, fewer than one in a million stars should be found at distances greater than $10\,R_h$. In practice, no members should be found at such great distances. 

Galactic tides  provide a possible mechanism for stars to populate the far outskirts of Scl and UMi \citep[e.g.][]{innanen+papp1979, walcher+2003}. 
However, there has been no convincing evidence for obvious tidal features---such as streams, tails, or shells---around either dwarf.
The aim of our study is to use N-body simulations to assess whether Galactic tides influence the outer regions of Scl and UMi and explain the presence of their far-outlying members. 

Accurate modelling of tidal interactions often requires idealized simulations that reach high numerical convergence and resolution. Cosmological simulations struggle to resolve tidal effects on dwarfs, which are 
often near the resolution limit and vulnerable to artificial disruption
\citep[e.g.,][]{PNM2008, vandenbosch+2018, santos-santos+2025}. To overcome
these challenges, idealized simulations typically model a single subhalo\footnote{A ``halo" is a gravitationally self-bound overdensity of dark matter. A subhalo is a halo that orbits inside the virial radius of a larger system. The virial radius is defined as the radius that encloses a mean overdensity equal to $200$ times the critical density for closure.} in an
analytic host potential, achieving excellent numerical resolution and convergence \citep[e.g.,][]{hayashi+2003, bullock+johnston2005, penarrubia+2010, EN2021}. 

A few recent studies have modeled the tidal evolution of Scl or UMi. Most relevantly, \citet{iorio+2019} have simulated Scl, also finding little tidal evolution. \citet{tchiorniy+genina2026} also modeled the tidal evolution of Scl, UMi, Fornax, Carina, but their main focus has been on how tidal effects may bias mass profiles recovered using Jeans' methods. No studies have yet fully explored tidal effects on Scl and UMi using updated structural and kinematic properties or accounted for orbital and tidal effects from the Large Magellanic Cloud (LMC), as we attempt to do in this study. 

Our paper is organized as follows.  Section~\ref{sec:gaia}  describes the density profiles of classical dSphs and compares them with those of Scl and UMi. Section~\ref{sec:methods} then describes our numerical N-body methods, initial conditions, and orbital integrations.  Section~\ref{sec:results} presents the results of our tidal simulations, while Section~\ref{sec:discussion} summarizes our main results and discusses our conclusions regarding the origin of extended stellar density profiles.


\section{dSph density profiles from \textit{Gaia}} \label{sec:gaia}

\begin{figure}
    \centering
    \includegraphics[scale=1]{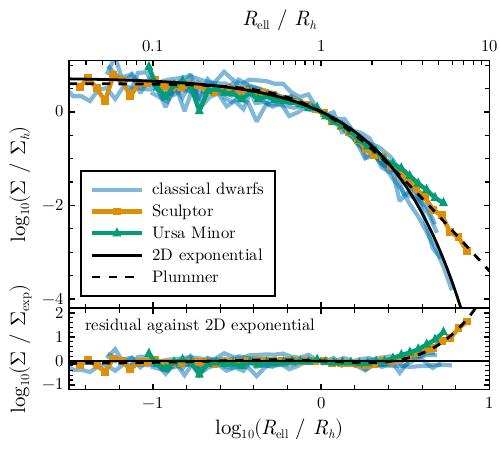}
    \caption{
        Top: The density profiles of
        Sculptor (orange squares), Ursa Minor (green triangles), and other
        classical dwarfs (blue lines, see text for the list of dSphs included)
        compared to 2D exponential (solid black
        line) and Plummer density profiles (dashed black line). 
        Density profiles are plotted as a function of the elliptical radius
        ($R_\text{ell}$, see equation~\ref{eq:r_ell}) scaled to the half-light
        radii ($R_h$) from \citet{MV2020a}, and surface density
        ($\Sigma$), scaled to the density at half-light radius $(\Sigma_h)$.
        Bottom: The residual density profiles when subtracting a 2D exponential.
        Note: We calculate the density profiles from \jj{} candidates with membership
        probabilities $P_\text{sat} > 0.2$.  We use bins of 0.05 dex
        width in $\log R$.
    }

    \label{fig:observed_density_profiles}
\end{figure}

Measuring the surface density profile of a resolved galaxy requires careful
consideration of whether any given star belongs to the system or not.
Without removing contaminating sources, faint
features may be hidden or of uncertain association. Fortunately, the data gathered by the \textit{Gaia} satellite now provides stellar parallaxes and proper motions, which greatly improve  membership assignment \citep{gaiacollaboration+2023}.

Here, we use the \citepos[hereafter \jj]{jensen+2024} \textit{Gaia} membership
probabilities. \jj{} used a Bayesian framework
incorporating proper motion, color--magnitude, and
spatial information to determine the probability that a given star
belongs to a satellite, $P_\text{sat}$. \jj{} extended
the algorithm presented in \citet{MV2020a, MV2020b} by additionally
including a secondary, extended spatial component.\footnote{
    We use
    the $P_\textrm{sat}$ values from the elliptical 2-component runs if a
    galaxy shows evidence for an outer component (i.e., Scl and UMi), or from the 1-component run
    otherwise. 
}
With this addition, \jj{} detected
candidate members out to ${\sim}10$ half-light radii ($R_h$) from the
centers of some galaxies. 

For our samples, we adopt a minimum probability of
$P_\text{sat} = 0.2$. Most stars have $P_\text{sat}$ values which are nearly 0 or 1, so the exact choice of probability threshold has little effect on the resulting sample. Even
at the probability threshold of 0.2, the total purity of line-of-sight velocity members remains
high (averaging ${\sim}90\%$, \jj{}).

Figure~\ref{fig:observed_density_profiles} compares the density profiles of Scl, UMi, and eight other classical dSphs. 
We include Fornax, Leo I, Leo II, Carina, Draco, Canes Venatici I, Sextans I, and Crater II. 
We exclude Sagittarius and Antlia II because of the high foreground density of Milky Way (MW) stars.
We use sphericalized elliptical radii for stars, given by
\begin{equation} \label{eq:r_ell}
    R_\text{ell} = \sqrt{a\,b},
\end{equation}
where $a$ and $b$ are the semi-major and semi-minor axes of an ellipse centred on the dwarf galaxy and coincident with the star.
The density profiles are shown out to the radius where the surface density reaches the background density of stars, which are selected by stars where the product of the probability distributions in proper motion and colour-magnitude spaces from \jj{} is higher for satellite stars than background stars. 

All classical dwarfs are well described by an exponential profile in the inner regions, as shown in Figure~\ref{fig:observed_density_profiles}.
In the outer regions, however, Scl and UMi deviate and show a clear  excess of stars over an exponential law. The deviation from an exponential grows outwards and  reaches over an order of magnitude at $6 R_h$. A Plummer profile better characterizes this excess (dashed black line, see equation~\ref{EqPlummer}), where the outer region of the profile approximates a power law rather than an exponential cutoff.


\section{Methods} \label{sec:methods}

This paper aims to assess whether the outer excess of stars seen in Scl and UMi may arise from Galactic tides. Since both galaxies are dark-matter dominated, we use N-body simulations, which assume that the evolution is well-represented by the evolution of collisionless particles within a single dark matter subhalo under a specified Milky Way potential. We choose an orbit consistent with the present-day locations of Scl and UMi while maximizing tidal effects. We shall also consider a model which includes a time-evolving potential motivated by the ongoing accretion of the Large Magellanic Cloud into the Galactic halo.  We provide below details on our assumed Galactic potentials (Section~\ref{milky-way-potential}), orbital modelling and initial conditions (Section~\ref{sec:point_orbits}), N-body methods (Section~\ref{sec:nbody_methods}), and the particle-tagging method used to track the evolution of the stellar component of the satellites (Section~\ref{sec:painting_stars}).

\subsection{Galactic Potential and Reference Frame \label{milky-way-potential}}

To convert observed positions and velocities to Galactocentric
coordinates, we use the Astropy v4 right-handed Cartesian Galactocentric frame
\citep{astropycollaboration+2022}. In this frame, $x$ is the direction that joins the Sun and the Galactic center, $y$ is the direction of the motion of the Local
Standard of Rest, and $z$ is the direction perpendicular to the
Galactic plane. The solar position is
$[-8.12, 0.00, 0.0208]\, \kpc$
\citep{gravitycollaboration+2018, bennett+bovy2019} and the solar
velocity is $\V_\sun = [-12.9, 245.6, 7.78]\ \kms$
\citep{reid+brunthaler2004, drimmel+poggio2018, gravitycollaboration+2018}.
The uncertainties in this reference frame are generally smaller than the
uncertainty in a dwarf galaxy's distance and/or tangential velocity.

For our MW-only models, we adopt the potential from \citet{EP2020}, which is
a re-parameterization of the \citet{mcmillan2011} Galactic potential.
The potential consists of a \citet{hernquist1990} bulge, a \citet{miyamoto+nagai1975} thin and thick disk, and a Navarro-Frenk-White \citep[][hereafter NFW]{NFW1996, NFW1997} dark matter halo.
The NFW density profile is defined as,
\begin{equation}\label{eq:nfw}{
\rho/\rho_s= \frac{1}{(r/r_s)(1+r/r_s)^2},
}\end{equation}
where $r_s$ and $\rho_s$ are the scaling radius and density, respectively. The NFW halo can also be parameterized in terms of $\vmax$ and $\rmax$, the maximum circular velocity of a halo and the radius where the circular velocity peaks.  
For the Milky Way, we adopt $\rmax = 43.7\,$kpc and $\vmax = 191\,\kms$. Equivalently, this corresponds to a dark matter halo with virial mass  $M_{200} = 1.266\times 10^{12}\,\Mo$ and $r_s=20.2\,\kpc$.

For models where the Galactic potential includes the presence of the Large Magellanic Cloud, we use the \texttt{L3M11} model from
\citet{vasiliev2024}. The \texttt{L3M11} potential is an evolving
multipole approximation of an N-body simulation including a live MW and
LMC dark matter halo. The potential includes a static MW bulge and disk,
evolving MW and LMC halos, and the MW reflex motion. In their
simulation, the MW was initially an NFW halo with $r_s=16.5\,$kpc and
$M_{200}= 9.84\times10^{11}\Mo$, and the LMC is modeled as an NFW halo with
$r_s=11.7$ and $M_{200} = 2.46 \times 10^{11} \Mo$. Note that the total
\texttt{L3M11} MW mass is slightly less massive than our initial \citet{EP2020}
MW-only potential.

\subsection{Orbits} \label{sec:point_orbits}

\begin{deluxetable*}{llllll}

\tablecaption{
Observed properties of Sculptor and Ursa Minor
\label{tbl:obs_props}
}

\tablehead{
\colhead{Property} 
& \colhead{Units} 
& \colhead{Sculptor} & \colhead{Reference} 
& \colhead{Ursa Minor} & \colhead{Reference}
}
\startdata
$\alpha$ & (deg)       & $15.0183 \pm 0.0012$ & 1 
                            & $ 227.2420 \pm 0.0045$ & 1\\
$\delta$ & (deg)       & $-33.7186 \pm 0.0007$ & 1
                            & $67.2221 \pm 0.0016$ & 1\\
distance modulus & (mag)           & $19.60 \pm 0.05$ & 2  
                            & $19.23 \pm 0.11$ & 5\\
distance & (kpc)              & $83.2 \pm 2$ & 2 
                            & $70.1 \pm 3.6$ & 5\\
$\mu_{\alpha*}$ & ($\masyr$)  & $0.099 \pm 0.002_\text{stat} \pm 0.017_\text{sys}$ & 3 
                            & $-0.124 \pm 0.004_\text{stat} \pm 0.017_\text{sys}$ & 3\\
$\mu_\delta$ & ($\masyr$)     & $-0.160 \pm 0.002_\text{stat} \pm 0.017_\text{sys}$  & 3
                            & $0.078 \pm 0.004_\text{stat} \pm 0.017_\text{sys}$ & 3\\
$\V_\text{los}$ & ($\kms$)    & $111.2 \pm 0.3\ $ & 4
                            & $-245.9 \pm 0.3_\text{stat} \pm 1_\text{sys}$ & 6\\
$\sigma_\V$ & ($\kms$)        & $9.7\pm0.2\ $ & 4
                            & $8.6 \pm 0.3$ & 6\\
$R_h$ & (arcmin)                 & $9.79 \pm 0.04$ & 1
                            & $11.62 \pm 0.1$ & 1\\
ellipticity                & \ldots & $0.37 \pm 0.01$ & 1
                            & $0.55 \pm 0.01$ & 1\\
position angle& (deg)   & $94\pm1$ & 1
                            & $50 \pm 1$ & 1\\
$M_V$ & (mag)                      & $-10.82\pm0.14$ & 1
                            & $-9.03 \pm 0.05$ & 1\\
$M_\star/L_\star$     &    \ldots  & $1.7$ & 7 &  $1.9$  & 7 \\
$M_\star$   & ($\Mo$)                & $3\times10^6$ & \ldots &  $7\times 10^5$  & \ldots
\enddata

\tablecomments{Rows are: right ascension ($\alpha$), declination ($\delta$), distance modulus, distance, absolute proper motion in right ascension ($\mu_{\alpha*}=\mu_\alpha\cos\delta$), proper motion in declination ($\mu_\delta$), line-of-sight systemic velocity ($\V_\text{los}$), line-of-sight velocity dispersion ($\sigma_\V$), sphericalized half-light radius ($R_h$), ellipticity, position angle, absolute V-band magnitude ($M_V$), stellar V-band mass-to-light ratio ($M_\star / L_\star$), and derived stellar mass ($M_\star$).
}

\tablerefs{(1) \citet[Sérsic fit]{munoz+2018}, (2) \citet[RR lyrae distance]{tran+2022}, (3) \citet{MV2020b}, (4) \citet{arroyo-polonio+2024}, (5) \citep{garofalo+2025}, (6) \citet{pace+2020}, average of their MMT and Keck results with systematic uncertainty based on the difference between the MMT and Keck means, (7) \citet{woo+courteau+dekel2008}.
}
\end{deluxetable*}

\begin{deluxetable}{llrrr}
\tablecaption{The present-day coordinates of our selected point-particle orbits with the $3\sigma$ smallest pericenter
}
\label{tbl:orbit_ics}

\tablehead{
\colhead{Property}
& \colhead{Units}
& \colhead{Scl: MW}
& \colhead{Scl: MW+LMC}
& \colhead{UMi: MW}
}

\startdata
distance &  (kpc)            & $82.6$    & $73.1$    & $64.6$\\
$\mu_{\alpha*}$ & ($\masyr$)    & $0.134$   & $0.137$   & $-0.158$\\
$\mu_\delta$ & ($\masyr$)       & $-0.198$  & $-0.156$  & $0.050$\\
$\V_\text{los}$ & ($\kms$)  & $111.2$   & $111.2$   & $-245.75$\\
pericenter & (kpc) & 43 & 39 & 30 \\
\enddata

\tablecomments{We assume the right ascension and declination from Table~\ref{tbl:obs_props}. }

\end{deluxetable}

\begin{deluxetable}{llrrr}[htp]
\tablecaption{The initial conditions for the N-body simulations presented in this work}
\label{tbl:nbody_ics}

\tablehead{
\colhead{Property}
& \colhead{Units}
& \colhead{Scl: MW}
& \colhead{Scl: MW+LMC}
& \colhead{UMi MW}
}
\startdata
$t_i$ & ($\Gyr$)        & $-9.17$   & $-2.0$    & $-9.67$\\
${x}_{i}$ & ($\kpc$)    & $-2.49$   & $4.30$    & $-17.40$\\
${y}_{i}$ & ($\kpc$)    & $-42.78$  & $138.89$  & $74.51$\\
${z}_{i}$ & ($\kpc$)    & $86.10$   & $125.88$  & $21.34$\\
$\V_{x,\,i}$ & ($\kms$) & $-20.56$  & $6.89$    & $14.27$\\
$\V_{y,\,i}$ & ($\kms$) & $-114.83$ & $-56.83$  & $48.62$\\
$\V_{z,\,i}$ & ($\kms$) & $-57.29$  & $52.09$   & $-114.08$\\
\hline
$\vmax$ & ($\kms$) & 31 & 25 & 38 \\
$\rmax$ & ($\kpc$) & 3.2 & 2.5 & 4.0 \\
$m_\text{DM}$ & ($\Mo$)  & 295 & 150 & 554\\
$R_{h,\,\text{exp}}$ & ($\kpc$) & 0.168 & 0.168 & 0.168 \\
$R_{h,\,\text{Plummer}}$ & ($ \kpc$) & 0.200 & 0.200 & 0.200 \\
\enddata

\tablecomments{
The first seven rows represent the initial time relative to today ($t_i$), position ($x_i, y_i, z_i$), and velocity ($\V_{x,\,i}, \V_{y,\,i}, \V_{z,\,i}$) of the simulation. The last four rows note the initial dwarf dark matter properties: the maximum circular velocity ($\vmax$), the radius of maximum circular velocity ($\rmax$), the dark matter particle mass $m_\text{DM}$, the initial half-light radius for the exponential stellar component ($R_{h,\,\text{exp}}$), and the initial half-light radius for the Plummer stellar component ($R_{h,\,\text{Plummer}}$). Scl and UMi have nearly the same physical size, so we adopt the same half-light radius for both galaxies.
}
\end{deluxetable}

To explore the possible orbits of Scl and UMi, we perform a Monte Carlo
sampling of their present-day positions and velocities from Table~\ref{tbl:obs_props}, including uncertainties. In practice, we draw 100,000 coordinates assuming Normal distributions without covariance. 
We then integrate each sampled position/velocity back in time for 10 Gyr using \agama{}
\citep{agama}. Since we find \citet{chandrasekhar1943} dynamical friction does not impact orbits substantially (not shown), we calculate point-mass particle orbits.

Tidal evolution depends mainly on pericentric distance. To maximize tidal effects, we select orbits with the smallest observationally-consistent pericenter for our N-body simulations.
We determine the final coordinates for each simulation by taking the median present-day coordinates of the sampled orbits with near the $3\sigma$ smallest pericenter\footnote{
    Specifically, the median of orbits with a pericenter less than the $2\times Q(3)\approx 0.0027$ quantile pericenter, where $Q(3)$ is the probability mass of a sample being greater than $3\sigma$ above the mean of a normal distribution.
}, resulting in values in Table~\ref{tbl:orbit_ics}. 
Given current observations, it is unlikely that either dwarf has had a significantly smaller pericenter than our chosen orbits.
We inject our N-body models into the MW potential at their first apocenter.

For the Scl MW+LMC model, we instead select for the smallest pericenter with respect to the LMC to test the possible influence of the LMC. We find similar results when choosing an orbit with the smallest MW pericenter instead of the smallest LMC pericenter. We also double Scl's present-day position and velocity errors when sampling point-particle orbits, as we do not account for uncertainties in the LMC trajectory. We only integrate this model backwards for $2\,\Gyr$ because, in the presence of the LMC, Scl is possibly completing its first orbit around the Milky Way and may have been outside the MW virial boundary before then (see Section~\ref{sec:results_orbits} below). 

The orbit of the UMi N-body model deviates from a point-particle orbit due to dynamical self-friction \citep[e.g.,][]{white1983}. To better match the present-day location, we iteratively adjust UMi's initial conditions using an action-angle based method presented in Appendix~\ref{sec:action_correction}, resulting in the position and velocity in Table~\ref{tbl:nbody_ics}. This method converges to an orbit agreeing
with the observed kinematics of UMi after 4 iterations.
Since Scl's orbit is less
strongly affected by tides, we do not carry out this correction for Scl.

\subsection{N-body Methods} \label{sec:nbody_methods}

We use \agama{} \citep{agama} to sample the initial N-body dark matter
halo of each dwarf, assuming the dark matter is well approximated by a spherical, isotropic NFW profile (equation~\ref{eq:nfw}). 
Since an NFW profile has a divergent total mass, we truncate the outer regions following
\begin{equation}\label{eq:trunc_nfw}
    \rho_\text{tNFW}(r) = e^{-(r/r_t)^3}\ \rho_\text{NFW}(r),
\end{equation}
where we set the truncation radius at $r_t = 20 \, r_s$. 

The stellar mass of both dwarf galaxies informs our choice of dark matter halo. We derive total stellar masses using the absolute magnitudes of \citet{munoz+2018} and the stellar mass-to-light ratios from \citet{woo+courteau+dekel2008}, which imply that Scl and UMi have stellar masses of $3\times10^6$ and $7\times10^5\,\Mo$, respectively.

We select cosmologically motivated values for the scaling parameters of the dark matter halos. 
Using the stellar mass--$\vmax$ relation from \citet{fattahi+2018} and the average mass-concentration relation from \citet{ludlow+2016}, we would expect that Scl and UMi's halos would presently have $\vmax \approx 31 \,\kms$, $\rmax\approx6\,\kpc$ and $\vmax \approx 27\,\kms$, $\rmax\approx5\,\kpc$, respectively (ignoring tidal evolution). 

Halos with these parameters, however, result in a stellar velocity dispersion smaller than observed for Scl and UMi after tidal evolution. We therefore adopt  $\vmax = 31\,\kms$ and $\rmax=3.2\,\kpc$ for Scl in the MW-only potential, $\vmax=25\,\kms$ and $\rmax=2.5\,\kpc$ for Scl in the MW+LMC potential, and
$\vmax=38\,\kms$ and $\rmax=4\,\kpc$ for UMi. 
Our initial dark matter halos are very near the mean concentration at $z=1$ from \citet{ludlow+2016}.

We assume that Scl and UMi's velocity dispersions arise from a virialized dark matter component. The crossing time at the half-light radius is $2\pi\, r_h / \mathrm{v}_\textrm{circ} \approx 2\pi\, r_h / \sqrt{3}\,\sigma_\V \approx 130\,\text{Myr}$. As the time of last pericentre is significantly greater ($\gtrsim400\,\text{Myr}$), we expect both systems to have since reached virial equilibrium in the inner regions.
Indeed, we find that the velocity dispersion in the simulations today is within one percent of the velocity dispersion calculated assuming virial equilibrium.\footnote{
    We derive the virial velocity dispersion using the Jeans equations from the
    empirical cumulative mass profile and stellar density profile of the final
    simulation snapshot (by integrating equation~\ref{eq:jeans}). For the
    non-equilibrium value, we compute the final velocity dispersion using the
    line-of-sight velocity of stars within three half-light radii of the
    center.
}
Binary stars may inflate the observed velocity dispersion of dwarf galaxies, especially for ultra-faint systems. The effect depends on the binary population, the availability of multi-epoch observations, and the intrinsic velocity dispersion of the system \citep{minor+2010, pianta+capuzzo-dolcetta+carraro2022, arroyo-polonio+battaglia+thomas2026}. In classical dwarf spheroidals, like Scl or UMi, binaries are unlikely to raise the velocity dispersion by more than ${\sim} 1\,\kms$  \citep[e.g.,][]{spencer+2017}. Scl and UMi's velocity dispersion likely accurately reflects the underlying dark matter component.

For simplicity, we assume an initially spherical galaxy. Both Scl and UMi have elliptical stellar components in projection (see Table~\ref{tbl:obs_props}) and may be highly elliptical in 3D \citep[e.g.,][]{an+koposov2022}. We explored an initially oblate N-body model (not shown) which evolved similarly to our spherical models and became more spherical over time \citep[see also][]{battaglia+sollima+nipoti2015}. A complete investigation into the tidal evolution of elliptical halos is beyond our scope.

We use \gadget{}, with an \agama{} external potential plug-in, for our N-body simulations. 
We adopt a Plummer-equivalent softening length of
\begin{equation}\label{eq:softening_length}
    \epsilon_\text{grav} = 0.014\,\text{kpc}\left(\frac{r_\text{max}}{6.0\,\text{kpc}}\right)\left(\frac{N}{10^7}\right)^{-1/2},
\end{equation} 
for a halo with $N$ particles. While our softening choice is about 3 times smaller than the values suggested by \citet{power+2003}, we find that this choice best balances convergence and computational efficiency for our simulations.
We use $10^7$ particles, a relative accuracy criterion of $\alpha=0.005$, an integration accuracy of $\eta=0.01$, and no minimum timestep.\footnote{i.e., tree nodes open if $M\,l/r^3 < \alpha |a|$ for node mass $M$, side-length $l$, distance $r$, and particle acceleration $a$; and particles must take time-steps smaller than $\sqrt{2\,\eta\,\epsilon_\text{grav} / |a|}$.}

To ensure that the initial conditions of the simulation are dynamically
relaxed, we run a halo first in isolation. We use the same isolation run for all models to save computational time, since gravity is scale-free. Our isolation halo was run for 5 Gyr assuming $\rmax=6\,\kpc$ and $\vmax=31\,\kms$, based on Scl's cosmologically-mean halo from above.
This model is converged down to a radius $r\approx 150\,\text{pc}$, where the enclosed mass changes by less than 2\% over $5\,\Gyr$ (see also the convergence radius from \citealt{power+2003}).

Next, we evolve the halo in the Galactic potential. We rescale and shift the
isolation-relaxed halo to our initial conditions (see Table~\ref{tbl:nbody_ics}). 
We then evolve the full N-body NFW model in the Galactic
potential until the present time, when the halo is closest to the present-day observed position of the selected dwarf in the MW
halo.\footnote{Specifically, the snapshot which best agrees (as measured
  by $\chi^2$) with the intended final position (in
  Table~\ref{tbl:orbit_ics}) and the observed position and velocity
  uncertainties. We use $R_h$ to determine the tolerance in $\alpha$ and
  $\delta$.}

We determine the halo center using a shrinking-spheres center method inspired by \citet{power+2003}.
First, we calculate the distance of all particles from the centroid, remove particles outside
the 95th percentile from the center, and recalculate the center of mass.
The procedure is repeated until the selection radius is less than
1\,kpc. 
For all subsequent time-steps, we
consider only particles retained from the previous iteration. 

\subsection{Sculptor and Ursa Minor's Initial Stellar
Components} \label{sec:painting_stars}

\begin{figure}
    \centering
    \includegraphics[scale=1]{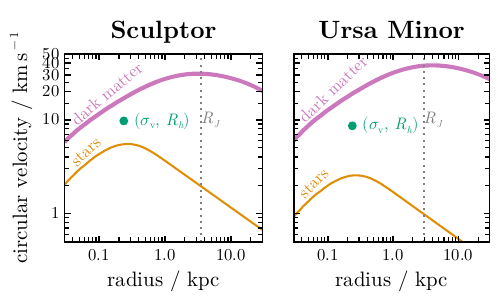}
    \caption{
       The initial circular velocity profiles as a function of radius
        for Sculptor and Ursa Minor for the MW-only models, assuming exponential stellar distributions. 
        The contribution of the stellar component is shown in orange; the dark matter circular velocity profile is shown in magenta. 
        We plot the models after 5\,Gyr of evolution in isolation, which are nearly identical to the initial conditions. 
        The vertical gray line marks the Jacobi radius (equation~\ref{eq:r_jacobi}), and the green dot marks the observed velocity dispersion and half-light radius of each dwarf at $z=0$ (the uncertainties in each quantity are smaller than the size of the dot).
    }
    \label{fig:initial_halos}
\end{figure}

To simulate the evolution of a stellar component, we ``paint" stars onto dark matter particles using the particle-tagging method \citep[e.g.,][]{bullock+johnston2005}. 
In essence, dark matter particles are assigned stellar weights based on the relative likelihood of finding a dark matter and stellar particle at each energy. The evolution of a collisionless stellar component is then emulated by weighting the dark matter particles by their stellar weights.
We briefly describe the procedure next, but refer interested readers to \citet{EP2020} for details.
For simplicity, we also assume that the galaxy is spherical and isotropic.

We assign a stellar weight ($P_\star$) to each particle using the ratio between the stellar and dark matter distribution functions\footnote{
    i.e., the phase-space density of particles. Note that since $f$ is formally
    defined in 6D phase space, there is an additional ``density of states''
    term in order to calculate the histogram of particles with ${\cal E}$,
    $dM/d{\cal E}$ \citep[section 4.4.5,][]{BT1987}.
}  ($f_\star$ and $f_\text{halo}$, respectively)
\begin{equation}
    P_\star({\cal E}) = \frac{f_\star({\cal E})}{f_\text{halo}({\cal E})},
\end{equation}
where $\cal{E}$ is the binding energy (defined as positive for bound particles).
We calculate the distribution function for both components via Eddington inversion, \begin{equation}{
    f({\cal E}) = \frac{1}{\sqrt{8}\, \pi^2}\left( \int_0^{\cal E} \frac{d^2\rho}{d\Psi^2} \frac{1}{\sqrt{{\cal E} - \Psi}}\ d\Psi + \frac{1}{\sqrt{\cal E}} \left(\frac{d\rho}{d\Psi}\right)_{\Psi=0} \right)
}
\end{equation} 
where $\rho$ is the stellar or dark matter density and $\Psi$ is the negative relative potential (normalized to vanish at infinity)  \citep[see eq. 4-140b in][]{BT1987}. 
We take the total potential $\Psi$ and dark matter density $\rho$ from the assumed analytic density profiles (equation~\ref{eq:trunc_nfw}) to reduce numerical noise. For $f_\star$, we derive the stellar density $\rho_\star$ from the stellar surface density $\Sigma_\star$ (specified below) using the inverse Abel transformation. We calculate the weights $P_\star$ for the snapshot saved after $1\,\Gyr$ of evolution in isolation, when the central region has relaxed but also before the end of the isolation run. 

For stellar components, we consider both a more compact exponential profile and an extended Plummer profile \citep{plummer1911}.  
A (projected) 2D-exponential profile is given by 
\begin{equation}\label{eq:exponential_law}
\Sigma_\text{exp} = \Sigma_0\exp(-\alpha\,R / R_h),
\end{equation}
with central surface density $\Sigma_0$, half-light radius $R_h$, and $\alpha \approx 1.6783$.
For a Plummer law, the density is,
\begin{equation}
\Sigma_\text{Pl} = \frac{\Sigma_0}{(1 + (R/R_h)^2)^2}. \label{EqPlummer}
\end{equation}
We adjust the initial half-light radii (in Table~\ref{tbl:nbody_ics}) such that the final simulated density profile best fits the observed density profiles.

Figure~\ref{fig:initial_halos} shows the initial circular velocity profiles of the dark matter and stars used in our simulations. 
Note that the stellar component is about $10\times$ more compact than the dark matter. The mass is dominated by dark matter across the galaxy, which confirms our assumption that  stars do not contribute to the gravitational potential.


\section{Results} \label{sec:results}

\subsection{Orbits of Sculptor and Ursa Minor} \label{sec:results_orbits}

\begin{figure*}
    \centering
    \includegraphics[scale=1]{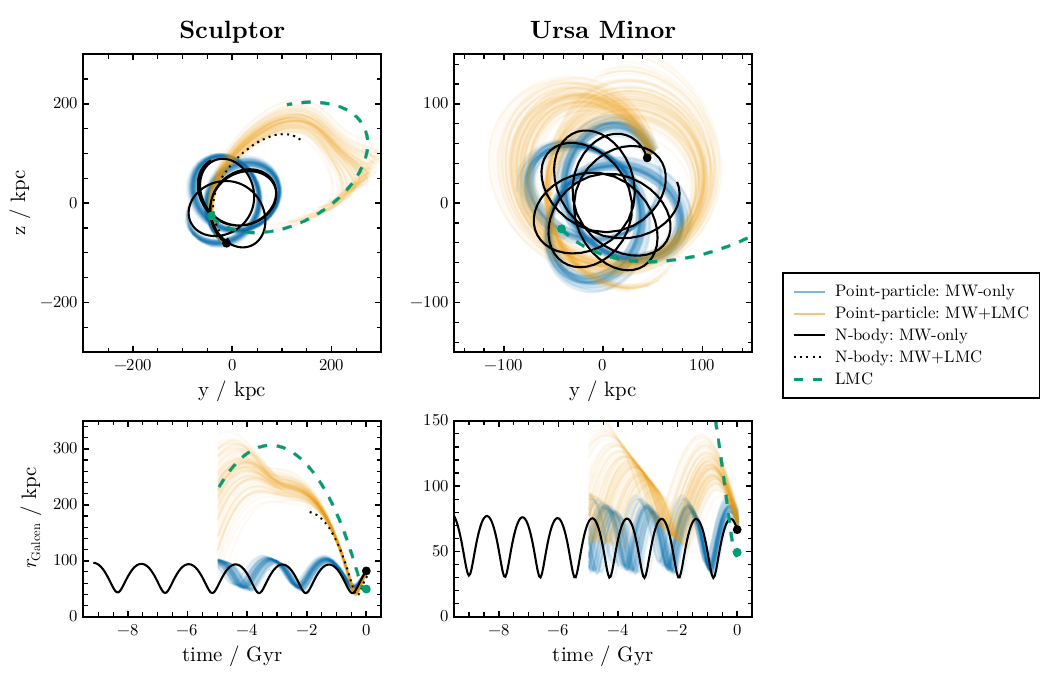}
    \caption{
        Point-mass orbits of Scl (left) and UMi (right) in the  MW-only potential (light blue lines) and the MW+LMC
        potential (light orange lines). Black lines show the orbits of our N-body runs, and
        the green dashed line shows the LMC trajectory. The top panels show orbits in the $y$--$z$ plane, and the bottom panels show orbits in terms of Galactocentric radius ($r_\text{Galcen}$) and time. We define the present as $\text{time}=0$, so negative values of time denote the past. 
}

    \label{fig:orbits}
\end{figure*}

Figure~\ref{fig:orbits} shows point-particle orbits of Scl and UMi in the MW-only and MW+LMC potentials. In the MW-only potential, both galaxies follow relatively well-constrained orbits consistent with long-term MW satellites. Scl experiences three pericentric passages over the past 5\,Gyr, reaching a minimum Galactocentric distance of $53\pm 3\,\kpc$. Over the same time, UMi experiences four smaller pericentric passages of Galactocentric radii $37\pm3\,\kpc$.  

Including the LMC dramatically modifies Scl's orbit. Scl is possibly coming into the Milky Way for the first time and recently completed its first pericentric passage. Scl has also passed close to the LMC about 100\,Myr ago, making the LMC potentially an important source of tidal stress. As we discuss later, the LMC-induced modification to Scl's orbit has the net effect of reducing the total tidal effect, mainly because it reduces the number of pericentric passages around the Milky Way.

In contrast, including the LMC increases UMi's pericenter to $\approx 58\,\kpc$ and lengthens the orbital period, reducing tidal forces overall.\footnote{Note that our MW-only model has a heavier MW mass than the MW+LMC model, but the qualitative effects on each galaxy's orbit are largely similar when using the same MW model in each case.} Regardless, the orbit is qualitatively similar to the MW-only case.

For comparison, the black lines in Figure~\ref{fig:orbits} show the orbits of our N-body models for Scl and UMi. We adopt these orbits as representatives of their tidal history, with the caveat that we have chosen the smallest possible pericenter in order to emphasize tidal effects. 

\subsection{Tidal Effects} \label{sec:tidal_effects}

\begin{figure}
    \centering
    \begin{interactive}{animation}{umi_animation.mp4}
    \includegraphics[width=\columnwidth]{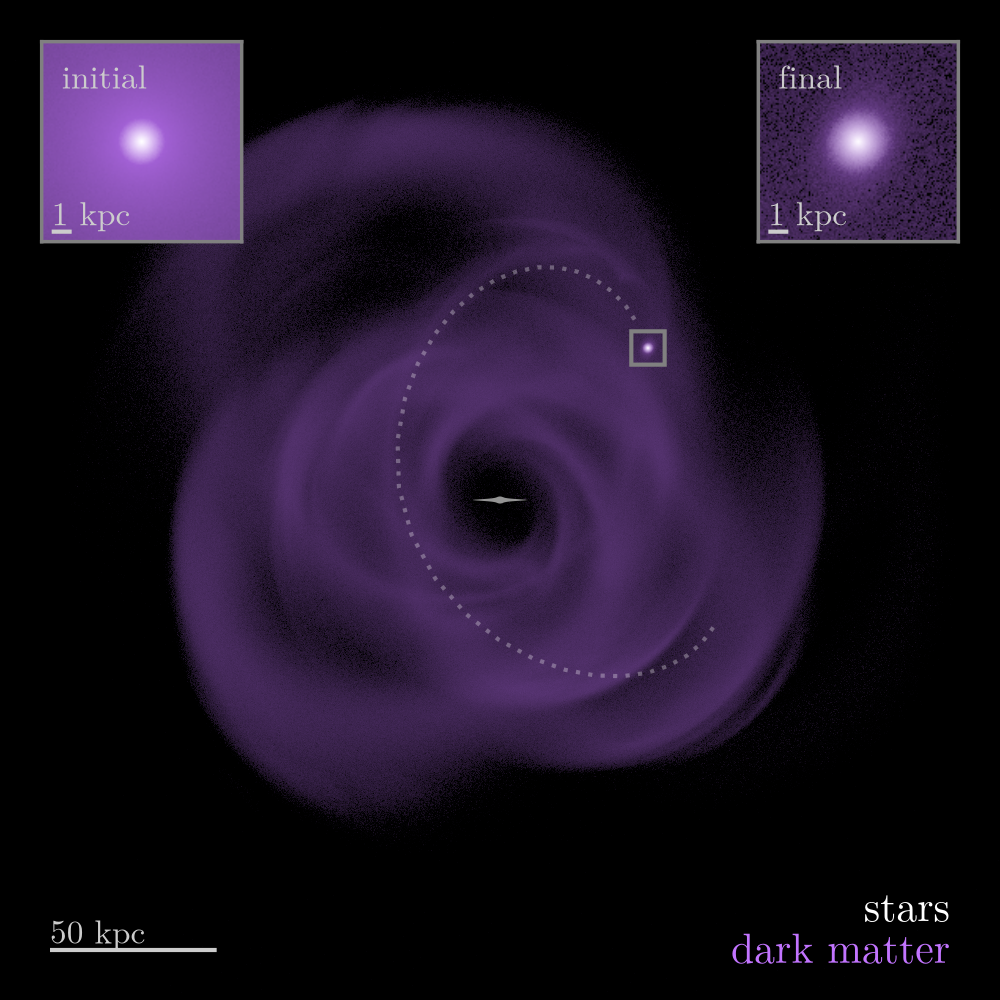}
    \end{interactive}
    \caption{
        The final dark matter and stellar distribution for the Ursa Minor MW-only model with exponential stars in the $y$--$z$ (near-orbital) plane. Dark matter is in purple with stars in white, each color range spanning 5 orders of magnitude from the peak value. The gray dotted line shows the past orbit up until the previous apocenter, and the insets zoom into the gray box in the main figure. 
        See Figures \ref{fig:scl_sim_images} and \ref{fig:scl_lmc_sim_images} for  similar images for the Scl N-body models. A ten-second animation illustrating the evolution from the initial time (9.7 Gyr ago) until the present day is available in the online article. 
        In the animated figure, UMi begins as a spherical dwarf galaxy at its apocentre. UMi then swoops towards the Milky Way, approaching its first pericentre, and releases substantial dark matter. With each subsequent pericentric passage (8 total), UMi is stripped of progressively more dark matter. The unbound dark matter forms diffuse streams along UMi's past and future orbit. However, UMi's stellar component remains bound and compact throughout the simulation. 
    }
    \label{fig:umi_sim_image}
\end{figure}

\begin{figure*}
    \centering
    \includegraphics[scale=1]{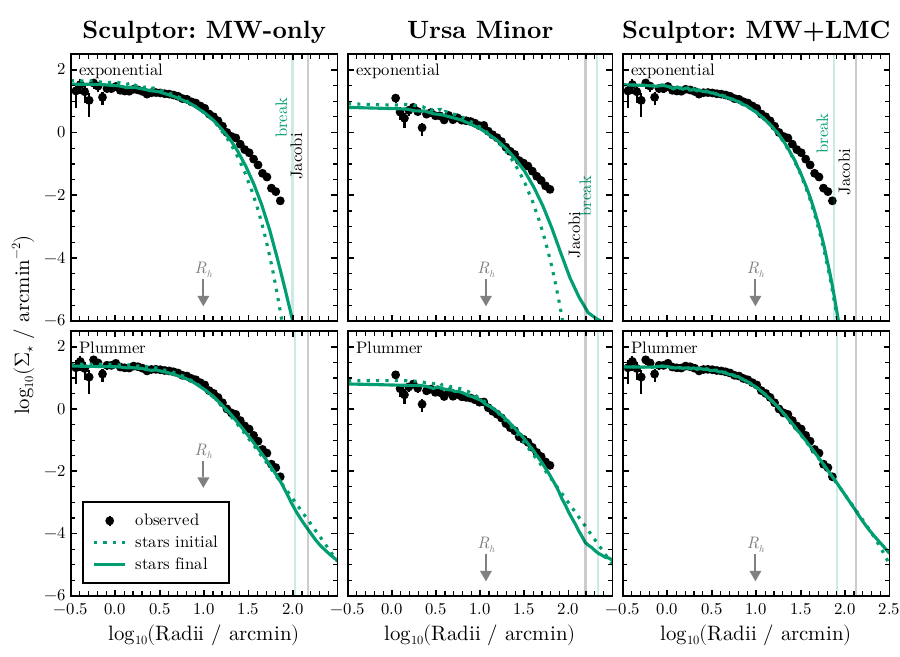}
    
    \caption{
        The stellar density profile of each model before (dotted) and after
        (solid) tidal evolution as compared to the observed density profile from \jj{} 
        (solid points). The top panels show 2D exponential initial stellar
        profiles, and the bottom panels show Plummer initial profiles. We mark the
        observed half-light ($R_h$) radius with arrows and the Jacobi and break
        radii with vertical lines. 
}
    \label{fig:stellar_model_density}
\end{figure*}

Next, we consider the tidal evolution of our N-body models in the MW-only case. Because of its smaller pericenter, UMi is more strongly affected by Galactic tides than Scl (also discussed below). Figure~\ref{fig:umi_sim_image} shows the final distribution of dark matter and stars for our N-body simulation of UMi. More than 95\% of UMi's initial dark matter mass has been  lost by the present day. 
However, the stellar component remains almost entirely bound (99.85\%) and spherical (ellipticity $< 0.02$).\footnote{
    We define ellipticity here as $e=1-c/a$ where $a, b, c$ are the semi-axes
    in descending order. We calculate the ellipticity from the eigenvalues
    of the reduced moment of inertia tensor
    \citep[e.g.,][]{dubinski+carlberg1991}. For bound particles, we assume
    spherical symmetry and iteratively remove unbound particles, recalculating
    the potential.
}
Over the simulation, the stellar component only expands slightly due to the reduction in the galaxy's inner mass (see insets of Figure~\ref{fig:umi_sim_image}). 

The left two panels of Figure~\ref{fig:stellar_model_density} show the initial and final stellar
density profiles of each galaxy in the MW-only potential compared to observations (shown as black circles). The top panels assume that the initial stellar density profile of each dwarf is exponential; the bottom panels assume a Plummer profile instead. The final profiles resemble the initial profiles over the full radial range observed in \jj{} (apart from a slight augmentation of the profile scale from the evolution of total mass). UMi's exponential stellar model experiences the strongest evolution, with the stars expanding slightly and flattening past 100 arcminutes. 

Whether the initial profiles are exponential or Plummer models, the tidal evolution of the stellar component is small. Both profiles remain nearly unchanged during
tidal evolution. To detect tidal disturbances, observations would  need to  search for features at least two magnitudes fainter (in surface density) than the limits reached at present ($\approx 0.01$ \textit{Gaia}-detected stars per arcmin$^2$). We conclude that the tidal effects are so slight that they are unlikely to be observable in the near future.

Does this conclusion change when accounting for the effect of the LMC? We only consider Scl in what follows, since  UMi's orbit primarily expands in radius when including the LMC, implying weaker tidal evolution.
The right panel of Figure~\ref{fig:stellar_model_density} shows the tidal effects on Scl in the LMC+MW potential.
Because of the reduced number of pericenters,
the stellar profiles evolve even less than in the MW-only case. Our conclusion is quite robust to reasonable changes in LMC mass and orbits. Our selected LMC model is already quite massive ($M_{200} \approx 3\times10^{11}\,\Mo$), and a numerical exploration of other possible LMC orbits (not shown) also supports our conclusion that accounting for the LMC potential only weakens the overall tidal effect on Scl.

\subsection{Characteristic Radii} \label{sec:break_radii}

Tidal effects imprint two characteristic radii on an orbiting satellite. One is
the pericentric \textbf{Jacobi radius}, $r_\text{J}$, which roughly delineates  the radius outside which stars should eventually become unbound.\footnote{ 
    While strictly valid only for circular orbits, assuming
$r_\text{peri}$ for the host-dwarf distance works as most stars are
lost during pericentric passages. Note that $r_\text{J}$ is also a 3D radius, as opposed to the break radius below.} $r_\text{J}$ is defined as the radius from the satellite's center where its mean density equals three times the mean density of the host at the pericenter of its orbit \citep[eq. 7-84]{BT1987}; i.e.,
\begin{equation}\label{eq:r_jacobi}
    3\, \bar \rho_\text{MW}(r_\text{peri}) \approx \bar \rho_\text{dwarf}(r_J),
\end{equation}
If $r_J$ is comparable to
the visible extent of a galaxy, we should expect to find clear signs of
tidal disturbance.

The other characteristic radius is the  \textbf{break radius}, $R_{\rm break}$, which delineates the extent of the dynamically-relaxed, self-bound remnant after a pericentric passage \citep{aguilar+white1986, navarro1989}. Outside this radius, stars are still radially mixing, leading to the presence of a transient and outward-moving ``bump" in the outer density profile. The position of the break radius depends on the velocity dispersion, $\sigma_v$, and the time elapsed
since pericenter, $\Delta t$, 
\begin{equation}\label{eq:r_break}
    R_\text{break} = C\,\sigma_{v}\,\Delta t,
\end{equation} 
where the empirical constant is $C \approx 0.55$  \citep{penarrubia+2009}.

Both the Jacobi and break radii are marked in Figure~\ref{fig:stellar_model_density}, and indicate that tidal effects should only  be apparent outside ${\sim}100$ arcminutes in either galaxy. 
These radii fall outside the range of the observed density profile. Given these estimates, it would be surprising to observe any tidal effects without substantially deeper data. We conclude from this discussion that tides have not substantially affected the stellar density profiles of either Scl or UMi on their present orbits.

In Appendix~\ref{sec:extra_nbody}, we also explored whether our conclusion changes when considering other modifications to our assumptions about the adopted structure of the dark matter halo in our models. We have tested cored, more extended, and anisotropic models, finding that these halos evolve similarly over time. As the observed velocity dispersion and observed half-light radii of Scl and UMi constrain the mean density interior to $R_h$, alternative dark matter structures still require initial conditions resulting in a similar final dark matter content. 

One key caveat is our assumption that the long-term orbits of UMi and Scl have not evolved dramatically. Our model neglects possible evolution that may arise from halo triaxiality, the presence of the Galactic bar, the evolution of the MW potential with time, as well as perturbations from substructure
\citep[e.g.,][]{dsouza+bell2022}. 
While the pericentre may evolve significantly over time, the most recent pericentre remains relatively well-determined \citep{santistevan+2024}. The last pericentric distance also primarily sets tidal effects near the break radius (see discussion above).
For simplicity, we assume Scl and UMi have not experienced pericentric passages much smaller than those explored here.

\subsection{Expected Tidal Effects on the Classical dSph Satellites of the Milky Way} \label{sec:classical_density_vs_peri}

\begin{figure}
    \centering
    \includegraphics[scale=1]{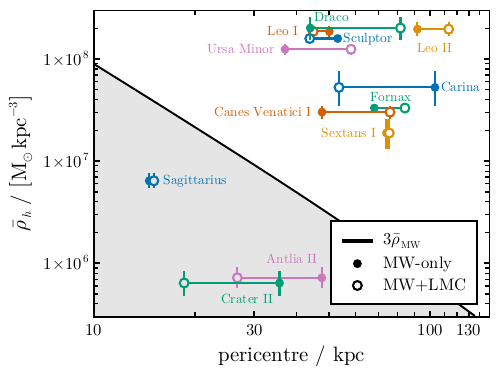}
    \caption{The mean density enclosed within the half-light radius ($\bar \rho_h$) versus pericentric distance for classical Milky Way dwarf spheroidal galaxies in a MW-only (filled circles) and MW+LMC potential (empty circles). The black line represents 3 times the enclosed density of the Milky Way, below which the Jacobi radius would be within a satellite's half-light radius, making it  vulnerable to disruption. Only Sagittarius, Crater II, and Antlia II fall in the region of likely tidal disruption. 
    }
    \label{fig:classical_jacobi}
\end{figure}

As discussed above, the lack of obvious tidal effects on the stellar components of Scl and UMi is not surprising given their high density compared with the host density at pericenter. We show this more explicitly in 
Figure~\ref{fig:classical_jacobi}, where we compare the mean inner (i.e., within $R_h$) density of all classical dSph satellites with the mean density of the Galaxy enclosed within the pericenter of their orbit (see also \citealt{pace+erkal+li2022}). Mean densities ($\bar\rho_h$) are computed using the \citet{wolf+2010} mass estimator. Pericenters are calculated from the mean orbits integrated backwards over 5 Gyr in our adopted MW and MW+LMC potentials using parameters from Appendix~\ref{sec:extra_pictures_and_obs}. We show two points for each satellite, corresponding to pericentric distances in the MW-only or MW+LMC potentials. 

It is clear from this plot that only Sagittarius, Antlia II, and Crater II fall within the region of likely tidal disruption. This is confirmed by the fact that Sagittarius has a prominent stream, and that recent observations demonstrate that Crater II and Antlia II show signs of tidal disruption \citep{ ji2025,vivas+2026}. The other classical dSphs, including Scl and UMi, have densities too high for tidal effects to reach the stellar component. Tides do not appear to  play a major role in the evolution of most MW classical dSphs.

\section{Discussion}

\subsection{Evidence for multiple populations} \label{sec:multiple_pops}

\begin{figure}
    \centering
    \includegraphics[scale=1]{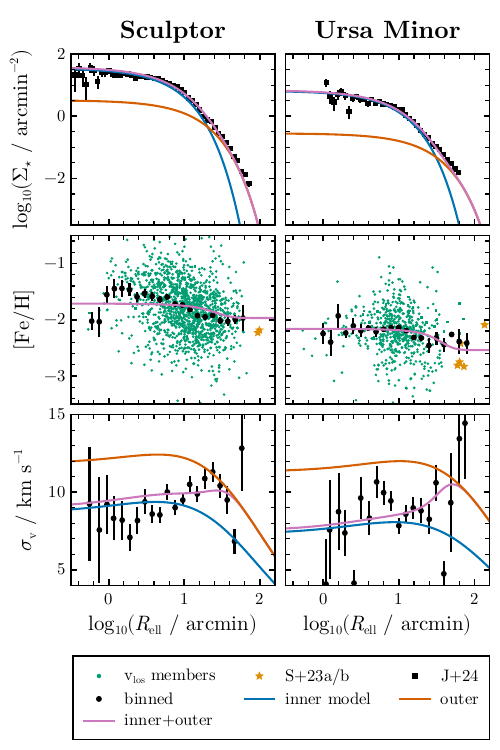}

    \caption{
        \textbf{Top:} Density profiles of Sculptor (top left) and Ursa Minor (top right) from \jj{} data (black squares) with our double exponential fits (see
        Appendix~\ref{sec:double_exp_fits}). Blue, rust, and magenta lines
        represent the inner, outer, and combined profiles, respectively.
        \textbf{Middle:} The metallicities of member stars in both galaxies as a function of 
        radius. For Sculptor, we plot stars from
        APOGEE, \citet{tolstoy+2023}, and
        \citet[][orange stars]{sestito+2023a}. For Ursa Minor, we
        plot stars from APOGEE, \citet{pace+2020}, and \citet[orange stars]{sestito+2023b}. Black points show the median metallicities of observed stars from the radial velocity sample in bins of $R_\textrm{ell}$. The magenta line
        represents a simple mixture model of metallicity, transitioning from
        higher to lower metallicity based on the relative inner and outer
        component densities (see Appendix~\ref{sec:double_exp_fits}). 
        \textbf{Bottom:} The velocity dispersion profiles of our binned radial velocity
        sample (black points) as compared to the velocity dispersion profile
        for exponential profiles (see text), with colors the same as the top panel.
    }
    \label{fig:metallicity_gradient}
\end{figure}

Since tides are insufficient to affect the outer profiles of  Scl and UMi, are there other hints as to why their density profiles differ from other classical dSphs? One clue  may be that Scl and UMi are known to host at least two distinct chemodynamic populations \citep[e.g.,][]{tolstoy+2004, battaglia+2008}.
The inner population is younger, higher metallicity, and dynamically colder, whereas the outer population is older, lower metallicity, and dynamically hotter.
Such distinct populations could indicate past accretion events or multiple generations of star formation, processes which may have had a role in the origin of their extended outer profiles. In this section, we explore whether Scl and UMi's multiple populations may be responsible for their unusual density profile.

The top row of Figure~\ref{fig:metallicity_gradient} shows Scl and UMi's density profiles decomposed into two distinct populations with different masses and radial extents. In Appendix~\ref{sec:double_exp_fits}, we describe our method to fit the \textit{Gaia} distribution of stars in Scl and UMi with a two-component exponential density profile (not including metallicity or kinematic information). In contrast to \citet{pace+2020, arroyo-polonio+2024}, we do not use metallicities or line-of-sight (los) velocities to constrain our two populations. The blue and rust curves in the top panels of the figure illustrate our best-fitting profiles, where the outer component dominates the inner one at $20$ arcmin and at $31$ arcmin from the center of Scl and UMi, respectively. 
However, without further constraints, many different decompositions into discrete populations are possible.

As an example, \citet{arroyo-polonio+2024} fit the populations of Scl using spatial positions, metallicities, and velocities of stars. They find Scl may be described by two chemodynamically distinct populations following Plummer profiles with (sphericalized) half-light radii of $6.2$ and $13$ arcminutes and with 66\% of stars in the outer population. We instead find two populations with half-light radii of 7.86 and 17.3 arcminutes, with only 31\% of stars in the outer population.  \citepos{arroyo-polonio+2024} decompose Scl into a somewhat different structure, but they use Plummer density profiles instead of the exponential profiles assumed here (see Appendix~\ref{sec:double_exp_fits} for a more detailed comparison). Scl may be best described with three separate components, as suggested in \citet{arroyo-polonio+2024}.

For UMi, \citet{pace+2020} find two populations with half-light radii of 11.4 and 23.2 arcminutes and with 24\% of stars in the outer component.\footnote{
 We show the results from \citet{pace+2020} for the Multiple Mirror Telescope/Hectochelle sample from \citealt{spencer+2018}. 
 \citet{pace+2020} also took new observations using the Keck telescope, which imply slightly different parameters that are in weaker agreement with our values, illustrating the large systemic uncertainties in isolating multiple populations within a dwarf galaxy.
}
Our two populations are similar: with half-light radii of 10.6 and 30 arcminutes, with 25\% of stars in the outer component. Therefore, UMi's extended density profile likely reflects this transition between an inner and outer stellar population.

When assuming the stellar density profile $\rho_\star$ and gravitational potential $\Phi$ of a galaxy, the velocity dispersion profile follows from  Jeans' equation for a spherical, isotropic system:
\begin{equation}\label{eq:jeans}
    \frac{d(\rho_\star \sigma_{\V,\,\textrm{los}}^2)}{dr} = - \rho_\star \frac{d\Phi}{dr},
\end{equation}
\citep[section 4.2]{BT1987}.
The resulting dispersion profiles are shown in the bottom panel of Figure~\ref{fig:metallicity_gradient}, assuming an NFW potential with
($\rmax=2.5\,\kpc$, $\vmax=25\,\kms$) for Scl and ($\rmax=5\,\kpc$, $\vmax=27\,\kms$) for UMi.
For our best-fit double-exponential density profile, the velocity dispersion of both Scl and UMi should increase in the outer regions, as the outer, dynamically hotter component starts to dominate. There is a hint that the observed velocity dispersion profile indeed increases with radius, but our suggestion should be confirmed by a more detailed study. While tidally-affected galaxies should also show an increasing velocity dispersion, the velocity profile only deviates from the equilibrium case outside the break radius. Thus, the velocity dispersion profiles currently observed should not be affected by transient tidal features. We simply assume a cosmologically-motivated NFW profile that matches the present-day velocity dispersion, similar to the argument in Section~\ref{sec:methods}. However, we do not account for the uncertain ellipticity, anisotropy, and 3D geometry of either system \citep[e.g.,][]{vitral+2024, vitral+2026}, which adds further uncertainty to this analysis.

Another observation that may be used to constrain the two-component model is the presence of a metallicity gradient, which would arise naturally if each component had different average metallicities. Assuming that Scl's two components have metallicities $\mathrm{[Fe/H]}=-1.69$ (inner) and $-1.97$ (outer) derived from the two-population fit in Appendix~\ref{sec:double_exp_fits}, we are able to approximately fit the observed mean metallicity gradient in Scl (left-middle panel in Figure~\ref{fig:metallicity_gradient}). The same is true for UMi, where we assume the inner and outer components have $\mathrm{[Fe/H]} = -2.14$ and $-2.54$ respectively. While our simplistic model accounts for the main trends of velocity and metallicity, our constraints are limited by the few spectroscopic members in the very outskirts of Scl and UMi. 

Recently, \citet{sato+2025b, sato+2025c} conducted an analysis of UMi with deep Subaru/Hyper Suprime-Cam photometry.  
They found extended structure in UMi that is most prominent along the minor axis, inconsistent with the orientation expected for tidal tails. \citet{sato+2025c} then derived the age--metallicity distribution from their photometry, confirming the presence of two populations of different ages, metallicities ($\mathrm{[Fe/H]}=-2.2$ and $-2.5$, similar to ours), spatial extent, and without internal age gradients. Both studies favour a merger explanation (we discuss other possibilities below).

The presence of multiple populations is not a unique feature of Scl and UMi. 
Within the MW satellite system, Carina \citep[e.g.,][]{battaglia+2012},
Fornax \citep[e.g.,][]{battaglia+2006},
and Sextans \citep[e.g.,][]{battaglia+2011}
also host several distinct populations of stars.
Why only Scl and UMi, and not other classical dwarfs, have density profiles which may reflect underlying stellar populations is unclear, and future investigation is warranted.


\subsection{Interpretation of extended profiles} \label{sec:discussion}

Since a tidal origin for the outer density excess in Scl and UMi is unlikely, we consider briefly other possible explanations next.

\paragraph{Episodic star formation} 
Star formation may quench and reignite in dwarfs, creating successive stellar generations with possibly different distributions. 
External star formation triggers include 
tidal compression \citep{mayer+2001a, dong+lin+murray2003}, 
collisions with gaseous filaments or dwarf galaxies \citep{genina+2019},
perturbations from dark halos \citep{starkenburg+helmi+sales2016},
or shocks with the MW corona \citep{wright+2019}.
Dwarfs with episodic star formation history should contain evidence of corresponding star formation bursts.

\paragraph{Major mergers} 
In a dwarf merger, stars from the lower-mass galaxy are preferentially dispersed, forming an extended stellar component and population gradient \citep{benitez-llambay+2016, deason+2022}. 
A halo may also form from many accreted ultra-faint dwarf galaxies \citep{ricotti+polisensky+cleland2022}. 
A few local dwarfs are suspected to have undergone a major merger, including Tucana II, Andromeda II, and Phoenix \citep{lokas+2014, fouquet+2017, tarumi+yoshida+frebel2021, cardona-barrero+2021, querci+2025}.
If a galaxy has accreted other systems, it should harbour multiple stellar populations with distinct origins.

\paragraph{Tidal preprocessing} 
Some dwarf galaxies may have been tidally ``preprocessed'' by a massive satellite like \textit{Gaia}-Sausage-Enceladus
\citep[e.g.,][]{santistevan+2023, riley+2025}. 
Tidal preprocessing redistributes already-present stellar populations and may mimic a stellar halo. 
Key evidence suggestive of preprocessing may include distant dwarf stars or stellar streams.

\paragraph{Dynamical heating} 
Old stars in dwarf galaxies may be dynamically hotter than younger stars. Dynamical heating processes including
stellar feedback \citep[e.g.,][]{stinson+2009, maxwell+2012}, 
sub-subhalo interactions \citep{penarrubia+2025}, 
or even fuzzy dark matter interference fringes \citep[e.g.,][]{el-zant+2020}. 
However, most of these processes should operate similarly across dwarf galaxies, failing to explain why Scl and UMi have unusual density profiles.

We cannot yet provide a definitive interpretation of Scl and UMi's differing density profiles as current data do not allow us to distinguish between (or rule out) some of the previous scenarios. Yet, these interpretations are useful to revisit as observations and modelling of MW dwarf satellites improve.


\section{Conclusions } \label{sec:conclusions}

We have used N-body simulations to explore whether tides from the Milky Way or LMC may explain the extended outer density profiles  of Scl and UMi. 
When adopting orbits with the most extreme ($3\sigma$) pericenters permitted by current observational constraints, we find that tides are unable to influence the stellar surface density profile of either galaxy. 
We also find that the LMC strongly perturbed the past orbit of Scl, but the LMC changes the orbits of Scl and UMi so as to weaken the overall tidal effect. 
We conclude that the observed density excess in Sculptor and Ursa Minor is not tidal in origin. 

Instead, the outer excess of Scl and UMi are likely innate features, reflecting a peculiarity of their formation histories. 
We showed that a two-component model fit to the density profile of Scl and UMi alone reproduces some features of their observed metallicity and velocity dispersion profiles. 
While a number of scenarios may in principle explain an extended outer stellar density (see the discussion in Section~\ref{sec:discussion}), future observations hold the promise to discriminate between different alternatives.  
Precise chemistry (e.g., 4MOST, \citealt{skuladottir+2023} or extremely large telescopes, \citealt{jagourel+2018}) can test if an extended stellar population was accreted and could constrain properties of merged satellites or the star formation history. 
Deep photometry (from, e.g., Euclid \citealt{euclidcollaboration+2025} or the Rubin Observatory \citealt{ivezic+2019}) and precise kinematics may find or rule out signs of dynamical disequilibrium and tidal tails and constrain the presence of tidally induced features around dwarf galaxies.
Finally, star formation histories may differentiate scenarios that rely on sudden variation in the star formation history, such as strong bursts. 
Both upcoming observatories and novel simulation methods will be instrumental in uncovering the clues to the formation histories of Sculptor and Ursa Minor encoded in their outer density profiles.


\begin{acknowledgements}

We thank the reviewer whose comments have improved the clarity of this work. 

We acknowledge and respect the \Lekwnen{} (Songhees and X\textsuperscript{w}seps\schwa{}m/
Esquimalt) Peoples on whose unceded territory the University of Victoria stands, and the
\Lekwnen{} and ${\underbar{\rm W}}$S\'ANE\'C Peoples whose historical relationships with the
land continue to this day.

JFN acknowledges the hospitality of Durham University, the Max-Planck Institute of Astrophysics, and the Donostia International Physics Centre during the completion of this work.

JJ acknowledges support of a Natural Sciences and Engineering Research Council of Canada Discovery Grant RGPIN2018-03853.

RE acknowledges support from the National Science Foundation (NSF) grant
AST-2206046. Support for program JWST-AR-02352.001-A was provided by
NASA through a grant from the Space Telescope Science Institute, which
is operated by the Association of Universities for Research in
Astronomy, Inc., under NASA contract NAS 5-03127. This material is based
upon work supported by the National Aeronautics and Space Administration
under Grant/Agreement No. 80NSSC24K0084 as part of the Roman Large Wide
Field Science program funded through ROSES call NNH22ZDA001N-ROMAN.

This work used the DiRAC@Durham facility managed by the Institute for Computational
Cosmology on behalf of the STFC DiRAC HPC Facility (\url{www.dirac.ac.uk}). The equipment
was funded by BEIS capital funding via STFC capital grants ST/K00042X/1, ST/P002293/1,
ST/R002371/1 and ST/S002502/1, Durham University and STFC operations grant ST/R000832/1.
DiRAC is part of the National e-Infrastructure.

Computations were performed on the Niagara/Trillium supercomputer \citep{ponce+2019} at the
SciNet HPC Consortium \citep{loken+2010}. SciNet is funded by Innovation, Science
and Economic Development Canada; the Digital Research Alliance of Canada; the Ontario
Research Fund: Research Excellence; and the University of Toronto.

This work has made use of data from the European Space Agency (ESA) mission Gaia
(\url{https://www.cosmos.esa.int/gaia}), processed by the Gaia Data Processing
and Analysis Consortium (DPAC),
\url{https://www.cosmos.esa.int/web/gaia/dpac/consortium}. Funding for the DPAC
has been provided by national institutions, in particular the institutions
participating in the Gaia Multilateral Agreement.

\software{
\gadget{} \citep{gadget4},
\agama{} \citep{agama},
\texttt{astropy} \citep{astropycollaboration+2013, astropycollaboration+2018, astropycollaboration+2022},
\texttt{Makie} \citep{makie},
\texttt{Turing} \citep{turing}.
}
\end{acknowledgements}


\appendix

\section{Correcting the initial conditions of N-body simulations} \label{sec:action_correction}

In this section, we outline our method to correct the initial position and velocity to attain a specified final position for an N-body simulation by using differences in action-angle coordinates.

Actions, $\mathbf{J}$, are integrals of motion which are also canonical momenta \citep[see, e.g.,][section 3.5]{BT2008}. The corresponding conjugate variables are the angles $\bm\theta$ which, together with $\mathbf J$, form a canonical coordinate system for a Hamiltonian $H$.
Angle coordinates are periodic---angle variables are equivalent up to the addition of integer multiples of $2\pi$.
While actions are invariant in time ($\dot{\mathbf J} = 0$), angle variables increase linearly with time : $\bm\theta(t) = \bm\theta(0) + \bm \Omega\,t$ where $\bm\Omega = \partial H / \partial \mathbf J$ for a Hamiltonian $H$. 

\agama{} provides utilities for converting between position-velocity space and action/angle space. To compute actions and angles, \agama{} uses the St\"ackel fudge, which assumes the potential is well-approximated by a St\"ackel potential \citep[see]{BT2008}. For the inverse problem, \agama{} converts actions-angles to position-velocity space with the ``torus mapping'' code from \citet{binney+mcmillan2016}. While neither transformation is exact, the precision is sufficient for our needs. 

To correct the initial positions, we make two approximations. First, a change to the initial actions $\mathbf J_i$ results in approximately the same change in the final, N-body evolved actions $\mathbf J_f$. Second, if the initial actions are unchanged ($\Delta \mathbf J_i = 0$),  then a change to the initial angles $\bm \theta_i$ results in a commensurate change to the final action angles $\bm \theta_f$. 

With these approximations, our procedure for adjusting the initial conditions is as follows:
\begin{enumerate}
    \item Compute the action and angles of the center of the N-body subhalo at the beginning and end of the simulation: $(\bm \theta_i, \mathbf J_i)$ and $(\bm \theta_f, \mathbf J_f)$. Determine the target final actions and angles, $(\bm \theta_t, \mathbf J_t)$. 
    \item While the final actions are significantly different from the target actions, adjust the initial actions based on the final deviation in actions: $\Delta \mathbf J_i = \gamma(\mathbf J_t - \mathbf J_f)$, where $\gamma > 0$ is a scale factor to aid convergence as needed. Holding $\bm \theta_i$ fixed, convert the initial actions and angles back into positions and velocities, and rerun the simulation.
    \item Finally, adjust the initial angles based on the final deviation in angles $\Delta \bm \theta_i = \gamma(\bm \theta_t - \bm \theta_f)$, holding $\mathbf J_i$ fixed. Rerun the simulation until convergence. 
\end{enumerate}

\begin{figure*}
    \centering
    \includegraphics[scale=1]{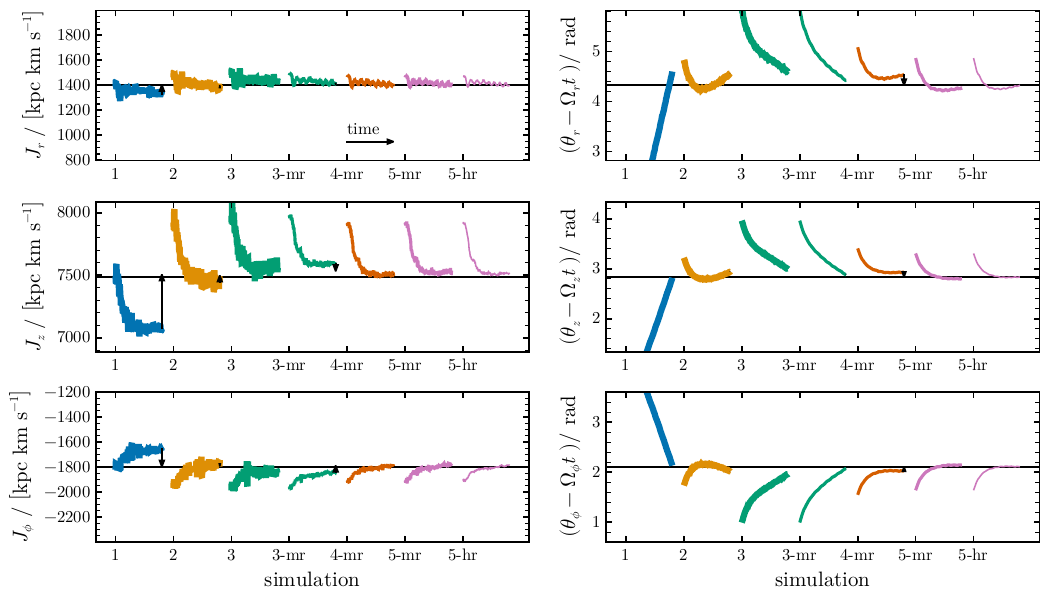}
    \caption{The evolution of initial actions (left) and relative angle variables (right) over time for each iteration of the simulation for Ursa Minor for $r$ (top row), $z$ (middle row), and $\phi$ (bottom row). 
    Each colored line shows the value of the action/angle over time from the initial conditions to the present day, scaled arbitrarily in the $x$-axis (representing time). The black arrows show the adjustments made to the initial conditions of the next simulations based on the final deviation. 
    The angle variables are relative to the expected evolution of the actions for the target point-particle orbit ($\theta_{a,\text{N-body}} - \theta_{a, \text{point particle}} =\theta_a -  \Omega_a\, t$).
    The suffixes -mr and -hr indicate that the model was run with $10^6$ or $10^7$ particles instead of $10^5$ (see also Table~\ref{tbl:action_shifts}).
    }
    \label{fig:action_corrections}
\end{figure*}

Figure~\ref{fig:action_corrections} illustrates this procedure as applied to our UMi simulation, and Table~\ref{tbl:action_shifts} records our action shifts and initial conditions for these simulations. As UMi mostly orbits in the $y$--$z$ plane, $J_z$ is the action with the largest magnitude. $J_z$ also evolves most strongly due to the dissipation of energy through dynamical self-friction. Additionally, actions fluctuate over time due to the limitations of the St\"ackel fudge and numerical noise from the simulation. 

We begin by using lower-resolution $10^5$ particle simulations (iteration 1). By shifting the initial actions by the final difference (black arrows), the next iteration achieves the target actions (iteration 2). For the third iteration, we increase the resolution (iteration 3 and 3-mr), which changes the evolution slightly and motivates one additional action correction.  
After this correction, the final actions are very near the expected values (simulation 4-mr). Finally, we adjust the angle variables and increase the resolution (simulation 5-hr), achieving the target position and velocities within uncertainties. 

Figure~\ref{fig:action_corrections} further tests our assumptions. Indeed, shifting the initial actions predominantly shifts the overall actions while the evolutionary shape of the actions remains similar.  As expected, shifts in the actions do change the evolution of the angles as the frequencies depend on the values of the actions. However, once the initial actions are fixed, changing the angle variables results in a nearly uniform shift to the angle evolution curves. For the case of UMi, the assumptions above appear sufficiently reasonable to achieve the intended final position and velocity.

While we have presented the above algorithm for one specific case, we have found success for simulations of other galaxies, so we anticipate this algorithm to work for satellites undergoing moderate to mild tidal disruption on halo orbits of axisymmetric potentials. Not shown here, we have also succeeded in extending this algorithm to mildly evolving potentials by using the initial axisymmetric potential to compute actions and angles, and choosing intermediate times of the simulation to compare and adjust the initial actions, extrapolating in time through point-particle orbits.

\begin{deluxetable*}{llrrrrrrr}
    \tablecaption{The changes to initial conditions for Ursa Minor resulting from our action-angle correction procedure 
    }

\tablehead{
\colhead{Parameter} &
\colhead{Units} &
\colhead{1} &
\colhead{2} &
\colhead{3} &
\colhead{3-mr} &
\colhead{4-mr} &
\colhead{5-mr} &
\colhead{5-hr}
}
    \startdata
    particles & \ldots & $10^5$ & $10^5$ & $10^5$ &  $\bm{10^6}$ & $10^6$  & $10^6$ & $\bm{10^7}$\\
    $\Delta J_r$ & ($\kpc\ \kms$)  & \ldots       &       $74.66$  &      $18.67$  &      \ldots     &     $-14.52$  &       \ldots       &       \ldots     \\
    $\Delta J_z$ & ($\kpc\ \kms$)  & \ldots       &       $431.39$  &      $51.85$  &      \ldots     &     $-62.22$  &       \ldots       &       \ldots     \\
 $\Delta J_\phi$ & ($\kpc\ \kms$)  & \ldots       &      $-143.11$  &     $-20.74$  &     \ldots      &      $51.85$  &       \ldots       &       \ldots   \\
               $\Delta \theta_r$  & (rad) &                 \ldots    &       \ldots & \ldots      &       \ldots      &       \ldots      &      $-0.22$  &       \ldots \\
               $\Delta \theta_z$  &      (rad) &             \ldots    &       \ldots & \ldots      &       \ldots      &       \ldots      &      $-0.10$  &       \ldots \\
            $\Delta \theta_\phi$  &        (rad) &           \ldots    &       \ldots & \ldots      &       \ldots      &       \ldots      &       $0.09$  &       \ldots \\
          $x_i$ & ($\kpc$)  &     $-16.36$  &     $-18.17$  &     $-18.36$  &    \ldots         &     $-18.07$  &     $-17.40$  &    \ldots         \\
          $y_i$ & ($\kpc$)  &      $70.19$  &      $74.99$  &      $75.74$  &    \ldots         &      $74.80$  &      $74.51$  &    \ldots         \\
          $z_i$ & ($\kpc$)  &      $20.19$  &      $20.73$  &      $20.89$  &    \ldots         &      $20.61$  &      $21.34$  &    \ldots         \\
   $\V_{x, i}$ & ($ \kms$)  &      $17.17$  &      $17.40$  &      $17.45$  &    \ldots         &      $17.13$  &      $14.28$  &    \ldots         \\
   $\V_{y, i}$ & ($ \kms$)  &      $36.21$  &      $35.00$  &      $34.86$  &    \ldots         &      $34.78$  &      $48.62$  &    \ldots         \\
   $\V_{z, i}$ & ($ \kms$)  &    $-118.09$  &    $-117.69$  &    $-117.41$  &    \ldots         &    $-117.66$  &    $-114.08$  &    \ldots         \\
  $\tilde\chi^2$  & \ldots &      $92$  &      $10$  &       $5.4$  &       $3.0$  &       $7.9$  &       $0.5$  &       $0.01$ 
     \enddata
     
    \label{tbl:action_shifts}

    \tablecomments{
    Rows: the number of particles used in the simulation,
    the shifts to actions and angles ($J_r$, $J_z$, $J_\phi$, $\theta_r$, $\theta_z$, $\theta_\phi$) for the initial conditions of the simulation (omitted if zero), 
    the initial positions and velocities (omitted if unchanged), and the final reduced $\chi^2$ of the final position of the simulation center. 
    $\tilde\chi^2$ is calculated by taking the difference in the final observed position of the simulation ($\alpha$, $\delta$, $\mu_{\alpha*}$, $\mu_\delta$, $\V_\text{los}$, and distance) as compared to the expected values (Table~\ref{tbl:orbit_ics}), using the uncertainties from Table~\ref{tbl:obs_props} and selecting $R_h$ as the uncertainty in $\alpha$ and $\delta$. We linearly interpolate the position and velocity of the center between simulation snapshots to stabilize $\tilde\chi^2$ . 
    }
\end{deluxetable*}

\section{Tidal evolution of subhalos with different dark matter structures} \label{sec:extra_nbody}

\begin{figure}
    \centering
    \includegraphics[scale=1]{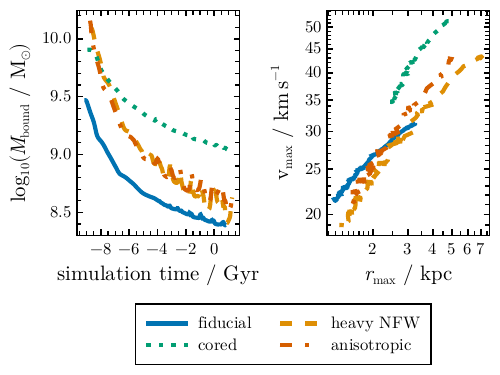}
    \caption{The tidal tracks for Scl models in the MW-only potential with alternate initial dark matter structure. The left panel shows the evolution of bound mass ($M_\text{bound}$) over time and the right panel shows $\vmax$ versus $\rmax$. We include the model from the main text (fiducial), a model with a heavier, more diffuse halo (heavy NFW), a model containing a dark matter core (cored), and a model with radial velocity anisotropy (anisotropic).}
    \label{fig:extra_tracks}
\end{figure}

Figure~\ref{fig:extra_tracks} shows the evolution of the total subhalo bound mass (left) and $\vmax$ versus $\rmax$ (right) for models of Scl with alternate halo structures. In all cases, we select parameters to be relatively near the expected cosmological distributions (see Section~\ref{sec:methods}) and match Scl's observed velocity dispersion after tidal evolution.
The heavier halo has $\vmax=43\,\kms$, $\rmax=7\,\kpc$, more than quadrupling the initial mass. We also explore a model with a dark matter core:
\begin{equation}
    \rho/\rho_s = \frac{1}{(1 + r/r_s)^2\ (r_c/r_s + r/r_s)},
\end{equation}
where $r_c$ is the core radius, and $r_s$ and $\rho_s$ are a scale radius and density, respectively \citet[e.g.,][]{errani+2023a}.  We set $M_s=(4\pi/3)\, \rho_s\,r_s^3=5.4\times 10^{9}\,\Mo$ and $r_s=r_c = 1.08\,\kpc$, using the same truncation as our NFW model (equation~\ref{eq:trunc_nfw}). The cored model illustrates an extreme case of a core and requires a significantly more massive halo to attain the observed velocity dispersion. 

We also consider a model with a radial velocity anisotropy. Velocity anisotropy is defined as $\beta = 1 - (\sigma_\phi^2 + \sigma_\theta^2) / 2\sigma_r^2$ where $\sigma^2$ is the velocity dispersion in the spherical $r$, $\phi$, or $\theta$ direction. We initialize a model with an Osipkov-Merrit velocity anisotropy profile, $\beta(r) = (\beta_0 + (r/r_a)^2) / (1 + (r/r_a)^2)$ rising from $\beta_0=0.2$ at the center to $\beta=1$ at infinity, with scale length $r_a= 4\,r_s$. We adopt $\vmax = 43\,\kms$ and $\rmax=5\,\kpc$ for the anisotropic model.

In each case, the tidal evolution is similar. The final halo reaches a similar mass to our fiducial model (excluding the cored model). Additionally, the final halo $\vmax$ and $\rmax$ are relatively similar to our fiducial model. The cored profile is an outlier but loses even less mass than our fiducial model since tidal evolution does not reach into the cored region. In any case, the detailed dark matter evolution may change depending on the structure of the initial halo, but the final structure and mass after tidal evolution are similar across cases here.

\section{Multiple population fits to density profiles} \label{sec:double_exp_fits}

\subsection{Spectroscopic Sample Selection} 
To create a sample of spectroscopic members, we compile line-of-sight velocities from
several surveys. For
Sculptor, we combine \citet{tolstoy+2023}; \citet{walker+2009};
\citet{sestito+2023a}; and APOGEE \citep[DR17,][]{abdurrouf+2022}. For
Ursa Minor, we combine \citet{spencer+2018}; \citet{pace+2020};
\citet{sestito+2023b}; and APOGEE. We then cross-match all catalogs to
\jj{} Gaia stars. If a study did not report \textit{Gaia} DR3 source IDs,
we match to the nearest star within 1--3 arcseconds. We combine
measurements of the same star using inverse-variance weighting. To
reduce likely binaries, we remove stars that are significant outliers in velocity.\footnote{Specifically, using that
  $\chi^2=\frac{s^2}{\delta \bar v^2}$, we remove stars with a
  $\chi^2$ larger than the 99.9th percentile of the $\chi^2$
  distribution with $N-1$ measurements.}

We build on \jj{}'s likelihood by adding multiplicative terms in the
total likelihood for the velocity consistency. We assume that the satellite and
background $v_\text{los, gsr}$ distributions are Gaussian in the
Galactic Standard of Rest (GSR, i.e., same location as ICRS but
velocities relative to the Galactic center). For the satellite, we adopt
a mean and standard deviation based on
Table~\ref{tbl:obs_props}, and, for the
background, mean $0\,\kms$ and dispersion
$\sigma_\text{halo} = 100\,\kms$ \citep[e.g.][]{brown+2010}. We select
stars with velocity-informed satellite membership probabilities of
greater than 0.2. For Scl, we find 1918 velocity member stars and 831 for UMi.

For UMi, we shifted the velocities of \citet{spencer+2018}
($-1.1\,\kms$) and \citet{pace+2020} ($+1.1\,\kms$) to account for
a systematic velocity offset. Otherwise, all studies appear to be on a
similar velocity scale.

We correct the velocities for the solar motion and the on-sky size of
the galaxy. We transform the velocities into the GSR and correct for the
apparent gradient induced by the dwarf's proper motion, assuming all stars are at the distance of the dwarf \citep[see,][]{WMO2008, strigari2010}. The
correction from both effects induces an apparent gradient of about
$1.3\,\mathrm{km\,s^{-1}\,deg^{-1}}$ for Sculptor and less for Ursa Minor. 

Using the spectroscopic samples, we derive the velocity dispersion profiles in the GSR frame through a Bayesian fit to stars within bins of 0.10 dex in $\log R_\textrm{ell}$. We solve for both the systemic velocity and velocity dispersion in each bin, accounting for the uncertainties of each individual star, using a No-U-Turn Sampler \citep[NUTS, an adaptive Hamiltonian Monte Carlo method][]{NUTS} as implemented in Turing \citep{turing}. For priors, we assume the systemic velocity is drawn from a normal distribution with mean 0 and standard deviation 100\,$\kms$, and the velocity dispersion is drawn from a uniform distribution between 0 and $20\,\kms$.

\subsection{Density Profile Fits} 

Here, we describe our fitting procedure to derive the double exponential fits and metallicity gradients in Figure~\ref{fig:metallicity_gradient}.

We extend \jj{}'s methodology to fit a double exponential with free inner and outer parameters. Specifically, we use a similar likelihood formulation except we hold the proper motion likelihood fixed, and allow the inner and outer density components to include free parameters for position angle, ellipticity, scale radius, and fraction of stars in the outer component ($f_\text{outer}$). We perform the fits using the No-U-Turn Sampler, assuming uniform priors for position angle, ellipticity, and $f_\text{outer}$, and log-uniform priors for scale radius.

To fit the metallicity gradient, we instead assume that the density profile follows the median double exponential fit above, and then fit a mixture model of the metallicity gradient as a sum of two Gaussian distributions. The four free parameters of this model are the mean and metallicity dispersion of the inner and outer components. Incorporating the structural uncertainties from the double exponential fit yields similar results, so we assume these parameters are fixed to their median value from above.

Table~\ref{tbl:double_exp_fits} contains the derived parameters for both Scl and UMi. We find strong evidence for multiple populations with different mean metallicities. Our outer population scale radius is smaller than in \jj{} for Scl but is similar for UMi.

\begin{figure}
    \centering
    \includegraphics[scale=1]{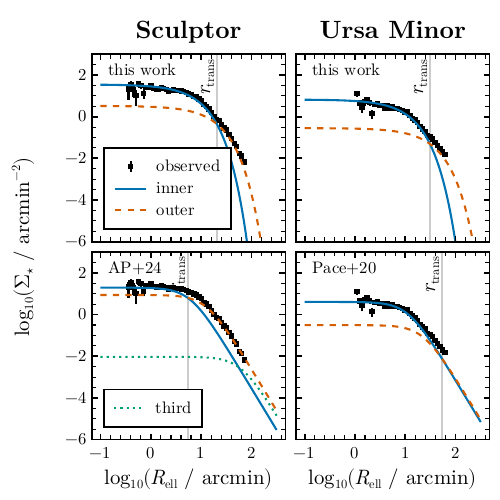}
    \caption{A comparison of the density decompositions in this work and the results from \citet{pace+2020} and \citet{arroyo-polonio+2024}. We plot the \jj{} density profiles (black squares), the inner component (blue solid lines), the outer component (orange dashed lines), the transition radius where the outer component begins to dominate (grey vertical line), and the third component of \citet{arroyo-polonio+2024} for Scl (green dotted line). The left and right panels show Scl and UMi, and the top and bottom panels show the results from this work and the literature, respectively.
    }
    \label{fig:decomposition_comparison}
\end{figure}

Finally, Figure~\ref{fig:decomposition_comparison} compares our best-fitting double-exponential density profile to the multiple chemodynamical populations of \citet{arroyo-polonio+2024} and \citet{pace+2020}. Promisingly, our two components are very similar to \citet{pace+2020}, except in the outer regions. As both \citet{arroyo-polonio+2024} and \citet{pace+2020} assume Plummer density profiles for each component, their outer structures are slightly different than ours. However, we do not find evidence of two Plummer populations based on the density profile alone.
As discussed in the main text, our decomposition does not directly agree with \citepos{arroyo-polonio+2024}. For Scl, they find more stellar mass in the outer (second) component, and both components have a more similar inner density. The transition between their two components occurs at a smaller radius of $R_\text{ell} \approx 5'$  compared to our transition location of $R_\text{ell} \approx 20'$. \citet{arroyo-polonio+2024} also tentatively detect a third population, but with an unconstrained scale radius.

\begin{deluxetable}{llll}
\tablecaption{The double exponential fit results for Scl and UMi}
\label{tbl:double_exp_fits}

\tablehead{
\colhead{Parameter} 
& \colhead{Units}
& \colhead{Inner} & 
\colhead{Outer}
}
\startdata
{\bf Sculptor}\\
$R_h$ & (arcmin)                & $7.86\pm0.21$        & $17.3\pm0.8$         \\
position angle & (deg)     & $98.8\pm2.7$          & $91.7\pm2.0$         \\
ellipticity  & \ldots                 & $0.244\pm0.022$     & $0.385\pm0.024$    \\
$f_\text{outer}$  & \ldots            & \ldots                           & $0.31\pm0.04$    \\
$R_\text{trans}$ & (arcmin)     & \ldots                           & $20.4\pm1.9$    \\
$\mu_\text{[Fe/H]}$ & (dex)     &  $-1.69 \pm 0.03$   &  $-1.97^{+0.06}_{-0.03}$  \\
$\sigma_\text{[Fe/H]}$ & (dex)  & $0.56^{+0.02}_{-0.14}$     & $0.35^{+0.31}_{-0.03}$    \\
\hline
\textbf{Ursa Minor} \\
$R_h$ & (arcmin)                & $10.57\pm0.33$       & $29.6\pm2.2$\\
position angle &  (deg)     & $50.4\pm1.1$          & $69^{+67}_{-38}$ \\
ellipticity  & \ldots                & $0.598\pm0.018$     &  $0.060^{+0.063}_{-0.042}$ \\
$f_\text{outer}$ & \ldots             & \ldots                           & $0.251\pm0.029$ \\
$R_\text{trans}$ & (arcmin)     & \ldots                           & $30.9\pm2.2$ \\
$\mu_\text{[Fe/H]}$ & (dex)     & $-2.144\pm0.020$      & $-2.54\pm0.09$ \\
$\sigma_\text{[Fe/H]}$ & (dex)  & $0.347\pm0.015$     & $0.49\pm0.07$ \\
\hline
\enddata

\tablecomments{Each column represents the median parameters with 16-84th percentiles for either the inner or outer exponential component. 
The parameters (in order) are the half-light radius ($R_h$), the position angle, the ellipticity, the fraction of stars belonging to the outer component $f_\text{outer}$, the radius where the outer component's density starts to dominate ($R_\text{trans}$), the mean [Fe/H] ($\mu_\text{[Fe/H]}$), and the standard deviation of [Fe/H] ($\sigma_\text{[Fe/H]}$).
}
\end{deluxetable}

\section{Additional plots and properties of other dwarf galaxies} \label{sec:extra_pictures_and_obs}

\begin{figure}
    \begin{interactive}{animation}{scl.mp4}
    \includegraphics[width=\columnwidth]{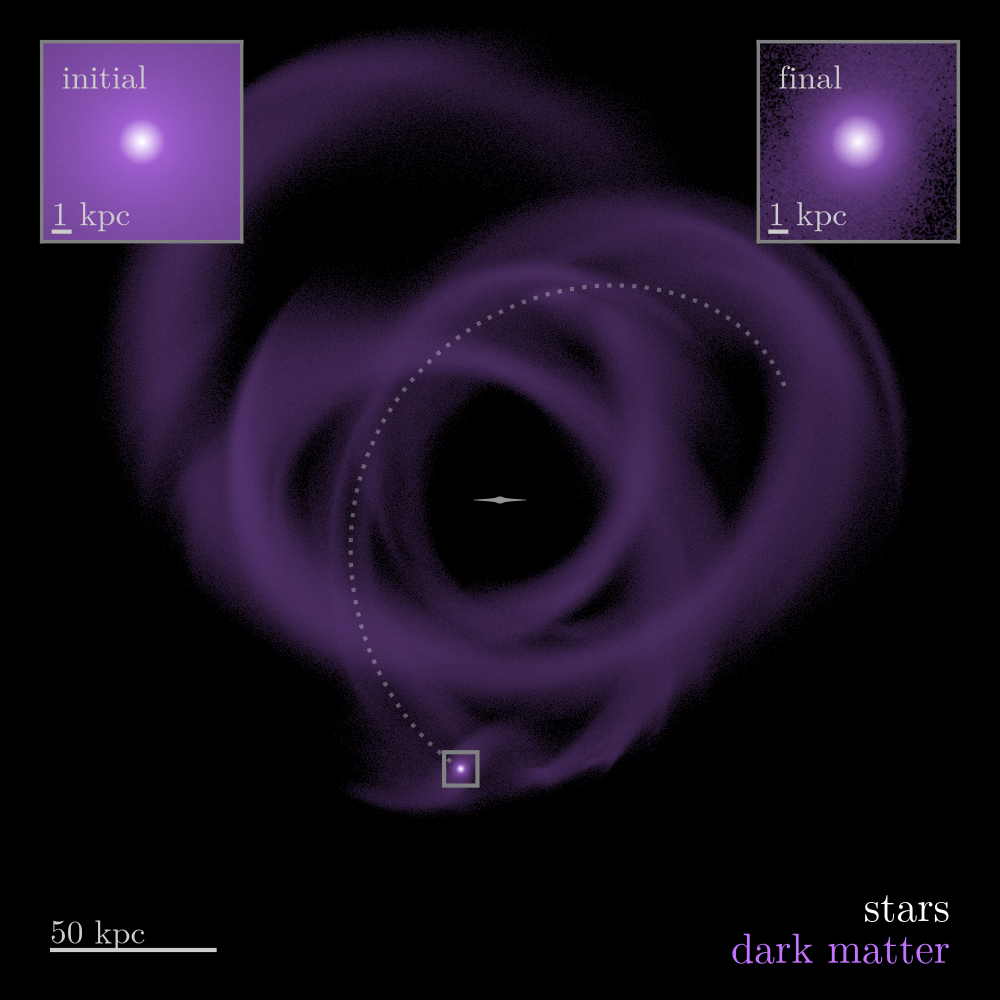}
    \end{interactive}
    \caption{
    Similar to Figure~\ref{fig:umi_sim_image} except for Scl in the MW-only potential. A ten-second animated version illustrating Scl's tidal evolution from 9.2 Gyr ago until today is available on the online journal. 
        In the animated figure, Scl begins at apocentre with a spherical dark matter halo. The galaxy is most strongly affected by its first pericentre. During the following five pericentres, Scl loses more mass. Scl appears less affected than UMi, with fainter evidence of dark matter streams.
    }
    \label{fig:scl_sim_images}
\end{figure}

\begin{figure}
    \begin{interactive}{animation}{scl_lmc_animation.mp4}
    \includegraphics[width=\columnwidth]{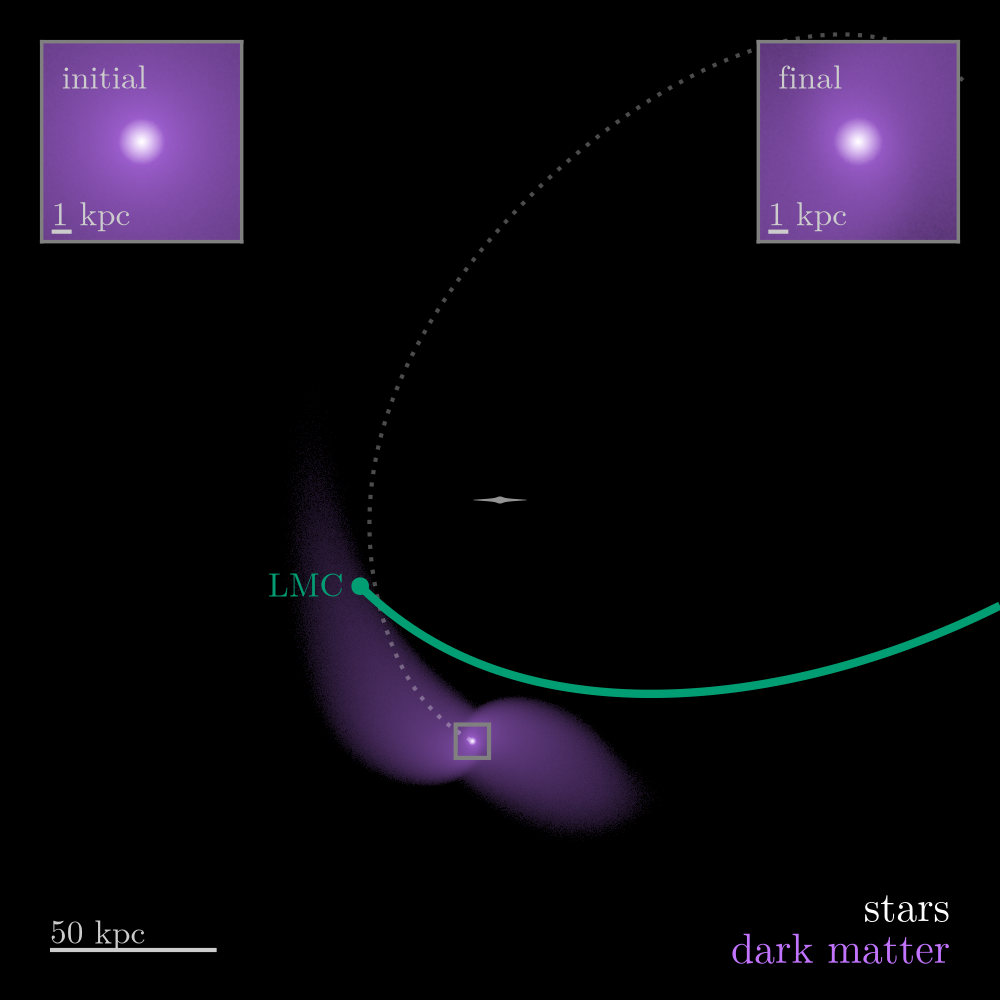}
    \end{interactive}
    \caption{
    Similar to Figure~\ref{fig:umi_sim_image} except for Scl in the MW-only potential. We also denote the path of the LMC with a green line.
    A seven-second animated version of this figure showing the 2 Gyr of simulated evolution is available on the online journal.
    The animated figure shows Scl and the LMC beginning far away from the MW centre. As Scl approaches the MW, Scl's halo elongates and forms slight dark matter tails. After passing its MW pericentre, Scl's dark matter ``stream"" begins to contract and appears unaffected by the near passage of the LMC. Scl evolves little compared to the MW-only model in Figure~\ref{fig:scl_sim_images}. 
    }
    \label{fig:scl_lmc_sim_images}
\end{figure}

Figures~\ref{fig:scl_sim_images} and \ref{fig:scl_lmc_sim_images} are similar to Figure~\ref{fig:umi_sim_image} except showing the final distribution of stars and dark matter for the Scl MW-only and MW+LMC models. As expected, Scl is less disrupted in the MW-only model and evolves little in the MW+LMC model.

Table~\ref{tbl:obs_props_other} contains the observed positions, kinematics, and distances of satellites besides Scl and UMi used for the orbits in Figure~\ref{fig:classical_jacobi}. Our parameters are primarily similar to \citet{MV2020a} for consistency, but we use updated kinematics and distances when available from \citepos{pace2025} Local Volume Database.

\begin{deluxetable*}{lrrrrrrrrl}
    \tablecaption{The properties of other MW classical satellites used for orbit integration in Figure~\ref{fig:classical_jacobi} 
 }
 \tablehead{
           \colhead{Galaxy}
         & \colhead{$\alpha$  }
         & \colhead{$\delta$  }
         & \colhead{DM}
         & \colhead{$\mu_{\alpha*}$}
         & \colhead{$\mu_{\delta}$ }
         & \colhead{$\V_\textrm{los}$  }
         & \colhead{$R_h$  }
         & \colhead{$\sigma_\V$  }%
         & \colhead{References } \\%
         & \colhead{(deg)}
         & \colhead{(deg)}
         & \colhead{(mag)}
         & \colhead{($\masyr$) }
         & \colhead{($\masyr$) }
         & \colhead{($\kms$) }
         & \colhead{(arcmin) }
         & \colhead{($\kms$)  }
 }
         
    \startdata
     Sagittarius &   $284.0951$ &  $ -30.5499$ & $ 17.10\pm0.15$ &      $-2.68$ &      $-1.39$ &     $143.90$ & $ 205\pm7$ & $ 12.91\pm0.45$ & 1, 2 \\
       Antlia II &   $143.8867$ &  $ -36.7672$ & $ 20.47\pm0.09$ &      $-0.09$ &       $0.12$ &     $290.70$ & $ 60.00\pm5.67$ & $ 5.98\pm0.36$ & 3, 4, 5 \\
          Fornax &    $39.9971$ &  $ -34.4492$ & $ 20.77\pm0.05$ &       $0.38$ &      $-0.36$ &      $55.30$ & $ 15.60\pm0.20$ & $ 12.10\pm0.20$ & 3, 6, 7, 8 \\
           Leo I &   $152.1171$ &  $ +12.3064$ & $ 22.06\pm0.08$ &      $-0.05$ &      $-0.11$ &     $282.50$ & $ 2.76\pm0.03$ & $ 9.20\pm0.40$ & 3, 9, 10 \\
          Leo II &   $168.3700$ &  $ +22.1517$ & $ 21.84\pm0.13$ &      $-0.14$ &      $-0.12$ &      $78.00$ & $ 2.39\pm0.03$ & $ 7.40\pm0.40$ & 3, 11 \\
       Sextans I &   $153.2625$ &   $ -1.6147$ & $ 19.67\pm0.10$ &      $-0.41$ &       $0.04$ &     $224.20$ & $ 22.41\pm0.97$ & $ 7.90\pm1.30$ & 3 \\
          Carina &   $100.4029$ &  $ -50.9661$ & $ 20.12\pm0.11$ &       $0.53$ &       $0.12$ &     $222.90$ & $ 9.07\pm0.10$ & $ 6.60\pm1.20$ & 3, 12 \\
           Draco &   $260.0517$ &  $ +57.9153$ & $ 19.56\pm0.04$ &       $0.04$ &      $-0.19$ &    $-291.00$ & $ 8.31\pm0.08$ & $ 9.10\pm1.20$ & 3, 13 \\
CVn I &   $202.0146$ &  $ +33.5558$ & $ 21.62\pm0.06$ &      $-0.11$ &      $-0.12$ &      $30.90$ & $ 6.95\pm0.31$ & $ 7.60\pm0.40$ & 3, 14 \\
       Crater II &   $177.3100$ &  $ -18.4131$ & $ 20.35\pm0.02$ &      $-0.07$ &      $-0.11$ &      $87.50$ & $ 30.40\pm2.60$ & $ 2.70\pm0.30$ & 3 \\
   \enddata
   
    \label{tbl:obs_props_other}

    \tablecomments{
        The columns are: right ascension, declination, distance modulus, proper motion in right ascension, proper motion in declination, line-of-sight velocity, sphericalized half-light radius, velocity dispersion, and the references used.
    }
    \tablerefs{
(1) \citet{mcconnachie2012},
(2) \citet{an+walker+pace2024},
(3) \citet{MV2020a, MV2020b},
(4) \citet{vivas+2022},
(5) \citet{ji+2021},
(6) \citet{oakes+2022},
(7) \citet{munoz+2018},
(8) \citet{WMO2009},
(9) \citet{stetson+2014},
(10) \citet{mateo+olszewski+walker2008},
(11) \citet{spencer+2017a},
(12) \citet{karczmarek+2015},
(13) \citet{bhardwaj+2024}, 
(14) \citet{kuehn+2008}.  
    }
\end{deluxetable*}

\newpage

\bibliographystyle{aasjournalv7.bst} 
\bibliography{paper.bib}

\end{document}